\documentclass[showkeys,superscriptaddress,floatfix,prd,10pt,aps]{revtex4-2}
\usepackage{graphicx,epstopdf}
\usepackage[caption=false]{subfig}
\usepackage{appendix}
\usepackage[T1]{fontenc}
\usepackage{lmodern}
\usepackage[dvipsnames,x11names]{xcolor}
\usepackage[colorlinks=true,linkcolor=NavyBlue,citecolor=ForestGreen,urlcolor=NavyBlue]{hyperref}
\usepackage[sort&compress]{natbib}
\usepackage{lipsum}
\usepackage{morefloats}
\usepackage[pdf]{pstricks}
\usepackage{amsmath,amsthm}
\usepackage{amssymb}
\usepackage{amsfonts}
\usepackage{rotating}
\usepackage{cancel}
\usepackage{mathtools}
\usepackage{bbm}
\usepackage{dsfont}
\usepackage{bbold}
\usepackage{multirow}
\usepackage{ulem}
\usepackage{physics}
\usepackage{orcidlink}
\usepackage{colortbl}
\usepackage{multirow}
\usepackage{tabularx}
\usepackage{rotating,tabularx}
\usepackage{threeparttable}
\usepackage{dcolumn}

\def\pslash{p\!\!\!\slash }
\def\qslash{q\!\!\!\slash }
\def\xslash{x\!\!\!\slash }
\def\eslash{\varepsilon\!\!\!\slash }

\def\vel{\left|}
\def\ver{\right|}

\begin{document}

\title{Magnetic dipole moments of the singly-heavy baryons with spin-$\frac{1}{2}$ and spin-$\frac{3}{2}$}

\author{Ula\c{s} \"{O}zdem\orcidlink{0000-0002-1907-2894}}%
\email[]{ulasozdem@aydin.edu.tr}
\affiliation{Health Services Vocational School of Higher Education, Istanbul Aydin University, Sefakoy-Kucukcekmece, 34295 Istanbul, T\"{u}rkiye}

%\date{\today}
 
\begin{abstract}
The electromagnetic characteristics of singly-heavy baryons at low energies are responsive to their internal composition, structural configuration, and the associated chiral dynamics of light diquarks.  To gain further insight, experimentalists are attempting to measure the magnetic and electric dipole moments of charm baryons at the LHC. In view of these developments, we conducted an extensive analysis of the magnetic dipole moments of both      $\rm{J^P}=\frac{1}{2}^+$ and $\rm{J^P}=\frac{3}{2}^+$ singly-heavy baryons by means of the QCD light-cone sum rules. Our findings have been compared with other phenomenological estimations that could prove a valuable supplementary resource for interpreting the singly-heavy baryon sector. To shed light on the internal structure of these baryons we study the contributions of the individual quark sectors to the magnetic dipole moments. It was observed that the magnetic dipole moments of the spin-$\frac{1}{2}$ sextet singly-heavy baryons are governed by the light quarks. Conversely, the role of the heavy quark is significantly enhanced for the spin-$\frac{1}{2}$ anti-triplet and spin-$\frac{3}{2}$ sextet singly-heavy baryons. The contribution of light and heavy quarks is observed to have an inverse relationship. The signs of the magnetic dipole moments demonstrate the interaction of the spin degrees of freedom of the quarks. The opposing signs of the light and heavy-quark magnetic dipole moments imply that the spins of these quarks are anti-aligned with respect to each other in the baryon. As a byproduct, the electric quadrupole and magnetic octupole moments of spin-$\frac{3}{2}$ singly-heavy baryons are also calculated.  We ascertained the existence of non-zero values for the electric quadrupole and magnetic octupole moments of these baryons, indicative of a non-spherical charge distribution.
\end{abstract}
%\keywords{Electromagnetic form factors, singly-heavy baryons, Magnetic dipole moments, QCD light-cone sum rules}

\maketitle

\section{Introduction}\label{motivation}

Recent experimental endeavors have yielded a plethora of novel hadronic states, markedly enhancing the domain of hadron physics over the past two decades. Among these findings, the singly-heavy baryons emerge as a distinctive and intriguing category. These heavy-light hadronic systems not only exhibit a remarkable prevalence but have also drawn substantial interest from the hadron physics community.  A singly-heavy baryon is constituted by a single heavy quark and two light quarks, exhibiting a high degree of adherence to the principles of heavy quark symmetry. This approach facilitates the investigation of specific properties by capitalizing on the observation that the mass of the heavy quark is considerably larger than the typical energy scale associated with strong interactions. The spectroscopy of singly-heavy baryons entails the analysis of their mass spectrum, decay behavior, and production processes. These investigations contribute to a deeper comprehension of the non-perturbative aspects of strong interactions. In recent years, there have been notable advancements in the experimental investigation of singly-heavy baryons~ \cite{ARGUS:1993vtm, CLEO:1994oxm, ARGUS:1997snv, E687:1993bax, CLEO:2000mbh, BaBar:2006itc, Belle:2006xni, LHCb:2017jym, Ammosov:1993pi, CLEO:1996czm, Belle:2004zjl, CLEO:1995amh, CLEO:1996zcj, E687:1998dwp, CLEO:2000ibb, CLEO:1999msf, LHCb:2020iby, BaBar:2007xtc, Belle:2006edu, BaBar:2007zjt, BaBar:2006pve, LHCb:2017uwr, LHCb:2012kxf, LHCb:2020lzx, LHCb:2019soc, LHCb:2018haf, CMS:2021rvl,  LHCb:2018vuc,  LHCb:2020xpu, LHCb:2020tqd}. In light of the tenuous evidence for the existence of some of these states and the lack of clarity regarding the quantum numbers, it is evident that further experimental research is necessary. Consequently, further investigation is required in both experimental and theoretical research~\cite{Cheng:2015iom, Klempt:2009pi, Chen:2016spr, Cheng:2021qpd, Meng:2022ozq, Chen:2022asf}.

The electromagnetic characteristics of singly-heavy baryons at low energies are responsive to their internal composition, structural configuration, and the associated chiral dynamics of light diquarks.  To gain further insight, experimentalists are attempting to measure the magnetic dipole moments of charm baryons at the Large Hadron Collider (LHC)~\cite{Aiola:2020yam, Neri:2024rjv, Baryshevsky:2016cul, Fomin:2017ltw, Bagli:2017foe, Fomin:2019wuw}. The experimental technique is grounded in a fixed-target experiment at the LHC, wherein high-energy charm baryons are generated and directed through a bent crystal, which deflects their trajectories within the detector acceptance. The high electric field between the crystal's atomic planes induces discernible spin precession in the channeled charm baryons. This experimental setup aims to measure the magnetic dipole moments of singly-charm baryons with a precision of better than 10$\%$ \cite{Akiba:2905467}. 
Inspired by this recent progress, we undertake a comprehensive investigation into the magnetic dipole moments of singly-heavy baryons within the framework of the QCD light-cone sum rule. Furthermore, the electric quadrupole and magnetic octupole moments of spin-3/2 singly-heavy baryons are also calculated. As is well known, the method used in this study, namely QCD light-cone sum rules, is an effective and adequate technique for investigating both static and dynamical properties of hadrons~\cite{Chernyak:1990ag, Braun:1988qv, Balitsky:1989ry}.   
The magnetic dipole moments and radiative decays of the singly-heavy baryons have been studied in other theoretical models,  including heavy baryon chiral perturbation theory (HB$\chi$PT) \cite{ Wang:2018gpl, Meng:2018gan, Wang:2018cre},  lattice QCD (LQCD) \cite{Can:2013tna, Can:2015exa, Bahtiyar:2015sga, Bahtiyar:2016dom, Can:2021ehb}, chiral constituent quark model ($\chi$CQM) \cite{Sharma:2010vv}, bag model (BM) \cite{Simonis:2018rld}, non-relativistic quark model (NRQM) \cite{Bernotas:2012nz}, relativistic three-quark model (RTQM) \cite{Faessler:2006ft}, covariant baryon chiral perturbation theory (B$\chi$PT) \cite{Shi:2018rhk, Liu:2018euh, Shi:2021kmm}, QCD light-cone sum rule (LCQSR) \cite{ Aliev:2008sk, Aliev:2015axa, Aliev:2001ig, Aliev:2008ay, Ozdem:2018uue, Ozdem:2019zis, Aliev:2004ju, Aliev:2009jt, Aliev:2014bma, Ozdem:2023cri,  Aliev:2016xvq, Wang:2009cd, Wang:2010xfj, Zhu:1998ih}, hyper central model (HCM) \cite{Patel:2007gx}, chiral quark soliton model ($\chi$QSM) \cite{Yang:2019tst,  Kim:2021xpp}, effective quark mass scheme  (EMS)\cite{Mohan:2022sxm, Hazra:2021lpa, Dhir:2013nka}, potential model (PM) \cite{Majethiya:2008fe},  Gaussian expansion method \cite{Peng:2024pyl}, and constituent quark model (CQM) \cite{Wang:2017kfr}. Further details may be found in Refs.~\cite{Cheng:2015iom, Klempt:2009pi, Chen:2016spr, Cheng:2021qpd, Meng:2022ozq, Chen:2022asf} and in the references cited in this paper. It should be noted that the magnetic dipole moments of singly-heavy baryons have previously been obtained within the framework of the QCD light-cone method~\cite{Aliev:2008sk, Aliev:2015axa, Aliev:2001ig, Aliev:2008ay}. In this study, we have conducted a more comprehensive analysis of all potential singly-heavy baryons with diverse interpolating currents.
Note that several potential interpolating currents can be formalized for singly-heavy baryons. Nevertheless, considering the QCD sum rules and the quantum numbers of the studied baryons, such as spin, parity, and isospin, the number of possible interpolating currents that can be formulated can be slightly restricted. The most effective way of understanding the structure and properties of a particle consists of assigning a structure appropriate for the task and then determining the corresponding characteristics. Theoretical insights into the nature and substructure of the hadron can be gained from the aforementioned calculations. To achieve this objective, it is imperative to identify suitable interpolating currents, comprised of quark fields that align with the valence quark content and quantum numbers of singly-heavy baryons. The formation of diquarks in color antitriplet is corroborated by the attractive interaction that is induced by the one-gluon exchange. It can be concluded from the results of the QCD sum rules that the most probable configurations are the scalar $(C\gamma_5)$ and axial-vector $(C\gamma_\mu)$~\cite{Wang:2010sh, Kleiv:2013dta}. In light of the aforementioned considerations, we have chosen to focus our analysis on scalar and axial-vector diquark structures.
 
 The rest of the paper is organized as follows. In Sec. \ref{formalism}, the methodology employed for the magnetic dipole moments of both spin-$\frac{1}{2}$ and spin-$\frac{3}{2}$ singly-heavy baryons, along with the analytical results obtained with the help of this methodology, is presented. In Sect.~\ref{numerical}, numerical analyses of the obtained analytical expressions are performed and compared with the existing results in the literature. The manuscript ends with a summary in Sec. \ref{summary}. 

\begin{widetext}
 
\section{Formalism for Magnetic dipole moments}\label{formalism}

\subsection{Formalism of the spin-$\frac{1}{2}$ singly-heavy baryons}

From the following calculation of the correlation functions in a weak external electromagnetic field, labeled $\gamma$, the QCD light-cone sum rules for the magnetic dipole moments of spin-$\frac{1}{2}$ singly-heavy baryons can be derived,
 \begin{align} \label{edmn01}
\Pi(p,q)&=i\int d^4x e^{ip \cdot x} \langle0|T\left\{\rm{J}(x)\bar{\rm{J}}(0)\right\}|0\rangle _\gamma \, , %\\
\end{align}
where $\mathrm{J}(x)$ is the interpolating currents of the corresponding baryons. 

Baryons with single-heavy quarks belong to $\bar 3_F$ (anti-triplet) or $6_F$ (sextet) flavor representation. The anti-triplet $\bar 3_F$ of baryons contain isosinglet $\Lambda_Q$, and isodoublet $\Xi_Q$. The sextet $6_F$ of baryons contain isotriplet $\Sigma_Q$, isodoublet $\Xi^\prime_Q$, and isosinglet $\Omega_Q$. The ground state heavy baryons are made from the heavy quark Q with spin-parity $\mathrm{J^P} = \frac{1}{2}^+$ and a light diquark system with spin-parity $ \mathrm{J^P} = 0^+$ ($\Lambda$-type) and $1^+$ ($\Sigma$-type) moving in a s-wave state relative to the heavy quark \cite{Korner:1994nh}. Based on this information, we choose $\Sigma$-type interpolating currents for sextet baryons (S) and $\Lambda$-type interpolating currents for anti-triplets (A) and give their explicit forms as follows:
\begin{align}
\rm{J}^{S}(x)& = \varepsilon^{abc}\big[q_1^{a^T}(x) C\gamma_\alpha q_2^b(x)\big] 
 \gamma^\alpha\gamma_5 Q^c(x) \, , \label{curpcs1}\\
%%%%%%%%%%%%%%%%%%%%%%%%%%%%%%%%%%%%%%%%%%%%%%%%%%%%%%%%%%%%%%%%%%%%%%%%%%%%%%%
%%%%%%%%%%%%%%%%%%%%%%%%%%%%%%%%%%%%%%%%%%%%%%%%%%%%%%%%%%%%%%%%%%%%%%%%%%%%%%%
\rm{J}^{A}(x)& =   \varepsilon^{abc} \big[q_1^{a^T}(x) C\gamma_5 q_2^b(x)\big] 
  Q^c(x) \, , \label{curpcs2}
%%%%%%%%%%%%%%%%%%%%%%%%%%%%%%%%%%%%%%%%%%%%%%%%%%%%%%%%%%%%%%%%%%%%%%%%%%%%%%%%%%%%%%%%%%%%%%%%%%%%%%%%%%%%%%%%%%%%%%%%%%%%%%%%%%%%%%%%%%%%%%%%%%%%%%%%%%%%%
\end{align}
where $a$, $b$, and $c$ being color indices; $Q= c$ and $b$-quark;  and the $C$ is the charge conjugation operator.  The quark content of the spin-$\frac{1}{2}$ singly-heavy baryons is presented in Table \ref{quarkcon}.
\begin{table}[htp]
	\addtolength{\tabcolsep}{10pt}
		\begin{center}
		\caption{The quark content of the spin-$\frac{1}{2}$ singly-heavy baryons.}
	\label{quarkcon}
\begin{tabular}{lccccccccc}
	   \hline\hline
	  % \\
  Quarks& $\Sigma_{c(b)}^{++(+)}$&$\Sigma_{c(b)}^{+(0)}$& $\Sigma_{c(b)}^{0(-)}$&
  $\Xi_{ c(b)}^{\prime +(0)}$& $\Xi_{ c(b)}^{\prime 0(-)}$& $\Omega_{c(b)}^{0(-)}$&
  $\Lambda_{c(b)}^{+(0)}$& $\Xi_{c(b)}^{+(0)}$& $\Xi_{c(b)}^{0(-)}$\\
%  \\
\hline\hline
%\\
$q_1$&  u & u & d & u & d & s & u & u & d\\
%\\
$q_2$&  u & d & d &s & s & s & d & s & s\\
%\\
	   \hline\hline
\end{tabular}
\end{center}
\end{table}

By following the procedure of the QCD light-cone sum rule, we will first formulate the correlation function as a variation of the physical parameters of the hadron, such as mass, form factor, residues, and so on. In the context of hadronic characterization, a complete set of baryons with identical quantum numbers as the interpolating currents $\mathrm{J}(x)$ is substituted into the correlation functions. The procedures mentioned above make it possible to write correlation functions in the following form
 \begin{align}\label{edmn02}
\Pi^{Had}(p,q)&=\frac{\langle0\mid \rm{J}(x) \mid
{\mathrm{B_Q}}(p, s) \rangle}{[p^{2}-m_{\mathrm{B_Q}}^{2}]}
\langle {\mathrm{B_Q}}(p, s)\mid
{\mathrm{B_Q}}(p+q, s)\rangle_\gamma 
\frac{\langle {\mathrm{B_Q}}(p+q, s)\mid
\bar{ \rm{J}}(0) \mid 0\rangle}{[(p+q)^{2}-m_{\mathrm{B_Q}}^{2}]}+ \cdots , 
\end{align}

The matrix elements seen in the above expression are the following.
%
%\begin{widetext}
%\begin{align}
\begin{align}
\label{edmn005}
\langle0\mid \rm{J}(x)\mid {\mathrm{B_Q}}(p, s)\rangle=&\lambda_{\mathrm{B_Q}} \gamma_5 \, \nu(p,s),\\%\label{edmn04}
%\\\nonumber\\
\langle {\mathrm{B_Q}}(p+q, s)\mid \rm{\bar J}(0)\mid 0\rangle=&\lambda_{\mathrm{B_Q}}  \, \bar \nu(p+q,s)\, \gamma_5,\\%\label{edmn004},\\
\langle {\mathrm{B_Q}}(p, s)\mid {\mathrm{B_Q}}(p+q, s)\rangle_\gamma &=\varepsilon^\mu\,\bar \nu(p, s)\bigg[\big[f_1(q^2)
+f_2(q^2)\big] \gamma_\mu +f_2(q^2)
\frac{(2p+q)_\mu}{2 m_{\mathrm{B_Q}}}\bigg]\,\nu(p+q, s), %\\
%\nonumber\\
%%%%%%%%%%%%%%%%%%%%%%%%%%%%%%%%%%
\end{align}
where  $\lambda_{\mathrm{B_Q}}$ and $ \nu(p+q,s)$ are the pole residue and spinors   of the $\mathrm{B_Q}$ baryons, respectively, and; $f_1(q^2)$ and $f_1(q^2)$ are the form factors of the corresponding transitions.

The following expressions for the hadronic description of the correlation functions for magnetic dipole moments are obtained from the above equations and some mathematical manipulations:
\begin{align}
\label{edmn050}
\Pi^{Had}(p,q)&=\frac{\lambda^2_{\mathrm{B_Q}}}{[(p+q)^2-m^2_{\mathrm{B_Q}}][p^2-m^2_{\mathrm{B_Q}}]}
  \bigg[\Big(f_1(q^2)+f_2(q^2)\Big)\Big(
  2 (\varepsilon . p) \pslash -
  m_{\mathrm{B_Q}}\,\eslash \pslash
  -m_{\mathrm{B_Q}}\,\eslash \qslash
  +\pslash\eslash\qslash
  \Big)+ \cdots \bigg],
\end{align}

%To compute the $F_M(Q^2)$ magnetic form factor for these singly-heavy baryons, it is necessary to express the form factors in terms of the previously defined $F_i(Q^2)$. These expressions are provided below:
%\begin{align}
%\label{edmn07}
%F_M(q^2) &= f_1(q^2) + f_2(q^2),
%\end{align}  
%where $\eta
%= -\frac{q^2}{4m^2_{{\mathrm{B_Q}}}}$.    
%The expressions mentioned above permit the determination of the electromagnetic form factors of these doubly-charmed states. However, as we work with a real photon ($q^2 =0$), we require expressions for the corresponding form factors regarding the magnetic dipole moment. The following equations provide the relevant details:
Using the above form factors, the magnetic dipole moment at the static limit, $q^2=0$, is defined as follows:
\begin{align}
\label{edmn08}
\mu_{\mathrm{B_Q}} &= \frac{ e}{2\, m_{\mathrm{B_Q}}} \,\big[f_1(0)+f_2(0)\big].
\end{align}
%with $F_{M}(0) = f_1(0)+f_2(0)$.

In the context of QCD parameters, the operator product expansion (OPE) is performed by first contracting the corresponding quark fields with Wick's theorem.  By following the procedures mentioned above for the QCD description, the correlation function is derived concerning the quark propagators and the distribution amplitudes (DAs) of the photon. The result of these procedures is achieved in the following way:
\begin{align}
\label{QCD1}
\Pi^{\rm{QCD}-S}(p,q)&= i\varepsilon^{abc} \varepsilon^{a^{\prime}b^{\prime}c^{\prime}}\, \int d^4x \, e^{ip\cdot x} 
 \langle 0\mid \Big\{ \mbox{Tr}\Big[\gamma_{\alpha} S_{q_2}^{bb^\prime}(x) \gamma_{\beta}  
  \widetilde S_{q_1}^{aa^\prime}(x)\Big] (\gamma^{\alpha}\gamma_5 S_{Q}^{cc^\prime}(x) \gamma_5  \gamma^{\beta}) 
 \nonumber\\
&  -
\mbox{Tr}\Big[\gamma_{\alpha} S_{q_2 q_1}^{ba^\prime}(x) \gamma_{\beta}  
  \widetilde S_{q_1 q_2}^{ab^\prime}(x)\Big]
  (\gamma^{\alpha}\gamma_5 S_{Q}^{cc^\prime}(x) \gamma_5  \gamma^{\beta})
  \Big\}
\mid 0 \rangle _\gamma \,,\\
%\nonumber\\
%\end{align}
%\begin{align}
\Pi^{\rm{QCD}-A}(p,q)&= i\varepsilon^{abc} \varepsilon^{a^{\prime}b^{\prime}c^{\prime}}\, \int d^4x \, e^{ip\cdot x} 
 \langle 0\mid \Big\{ \mbox{Tr}\Big[\gamma_{5} S_{q_2}^{bb^\prime}(x) \gamma_{5}  
  \widetilde S_{q_1}^{aa^\prime}(x)\Big]S_{Q}^{cc^\prime}(x)  \nonumber\\
&-
\mbox{Tr}\Big[\gamma_{5} S_{q_2 q_1}^{ba^\prime}(x) \gamma_{5}  
  \widetilde S_{q_1 q_2}^{ab^\prime}(x)\Big] S_{Q}^{cc^\prime}(x)
  \Big\} 
\mid 0 \rangle _\gamma \,,
\label{QCD2}
\end{align}
where   
%\begin{equation*}
$\widetilde{S}_{Q(q)}^{ij}(x)=CS_{Q(q)}^{ij\mathrm{T}}(x)C$. 
Here 
\begin{eqnarray}
\widetilde{S}^{ab^\prime}_{q_{1}q_{2}}(x) = \left\{ \begin{array}{rl}
\widetilde{S}^{ab^\prime}_{q_{1}}(x), &\mbox{if}~q_1= q_2\\
0, &\mbox{otherwise},
\end{array} \nonumber \right. 
\end{eqnarray}

\begin{eqnarray}
S^{ba^\prime}_{q_{2}q_{1}}(x) = \left\{ \begin{array}{rl}
S^{ba^\prime}_{q_{2}}(x), &\mbox{if}~q_1 = q_2\\
0, &\mbox{otherwise}.
\end{array} \nonumber \right. 
\end{eqnarray}

%$S^{ab^\prime}_{q_{1}q_{2}}(x)$ and $S^{ba^\prime}_{q_{2}q_{1}}(x)$ exist when $q_1=q_2$ but it vanishes when $q_1\neq q_2$. 
The relevant light ($S_{q}(x)$) and heavy ($S_{Q}(x)$) quark propagators are written as~\cite{Yang:1993bp, Belyaev:1985wza}:
\begin{align}
\label{edmn13}
S_{q}(x)&= S_q^{free}(x) 
- \frac{\langle \bar qq \rangle }{12} \Big(1-i\frac{m_{q} \xslash}{4}   \Big)
%\nonumber\\
%&
- \frac{ \langle \bar qq \rangle }{192}
m_0^2 x^2  \Big(1 %\nonumber\\
%&
  -i\frac{m_{q} \xslash}{6}   \Big)
+\frac {i g_s~G^{\mu \nu} (x)}{32 \pi^2 x^2} 
%\nonumber\\
%& \times 
\bigg[\rlap/{x} 
\sigma_{\mu \nu} +  \sigma_{\mu \nu} \rlap/{x}
 \bigg],\\
%\nonumber\\
%\end{align}%
%and
%
%\begin{align}
S_{Q}(x)&=S_Q^{free}(x)
%\nonumber\\
%&
-\frac{m_{Q}\,g_{s}\, G^{\mu \nu}(x)}{32\pi ^{2}} \bigg[ (\sigma _{\mu \nu }{\xslash}
 % \nonumber\\
%&
+{\xslash}\sigma _{\mu \nu }) 
    \frac{K_{1}\big( m_{Q}\sqrt{-x^{2}}\big) }{\sqrt{-x^{2}}}
   %\nonumber\\
  %&
 +2\sigma_{\mu \nu }K_{0}\big( m_{Q}\sqrt{-x^{2}}\big)\bigg],
 \label{edmn14}
\end{align}%
with  
\begin{align}
 S_q^{free}(x)&=\frac{1}{2 \pi x^2}\Big(i \frac{\xslash}{x^2}- \frac{m_q}{2}\Big),\\
 %\nonumber\\
 S_Q^{free}(x)&=\frac{m_{Q}^{2}}{4 \pi^{2}} \bigg[ \frac{K_{1}\big(m_{Q}\sqrt{-x^{2}}\big) }{\sqrt{-x^{2}}}
+i\frac{{\xslash}~K_{2}\big( m_{Q}\sqrt{-x^{2}}\big)}
{(\sqrt{-x^{2}})^{2}}\bigg],
\end{align}
where $m_0$ is defined through the quark-gluon mixed condensate $ m_0^2= \langle 0 \mid \bar  q\, g_s\, \sigma_{\mu\nu}\, G^{\mu\nu}\, q \mid 0 \rangle / \langle \bar qq \rangle $, $G^{\mu\nu}$ is the gluon field-strength tensor, and $K_n$'s being the modified second type Bessel functions.  Here, we use the following integral representation
 of the  modified second-type Bessel function,     
\begin{equation}\label{b2}
K_n(m_Q\sqrt{-x^2})=\frac{\Gamma(n+ 1/2)~2^n}{m_Q^n \,\sqrt{\pi}}\int_0^\infty dt~\cos(m_Qt)\frac{(\sqrt{-x^2})^n}{(t^2-x^2)^{n+1/2}}.
\end{equation}

Two different contributions to the correlation function can be observed in Eqs.~ (\ref{QCD1})-(\ref{QCD2}). The first one is the perturbative component, which occurs when a photon is radiated at short distances. The second one is the non-perturbative component, which arises when a photon is radiated at long distances. It is useful to use the following equation to calculate contributions from short distances, 
\begin{align}
\label{free}
S_{q(Q)}^{free}(x) \rightarrow \int d^4y\, S_{q(Q)}^{free} (x-z)\,\rlap/{\!A}(z)\, S_{q(Q)}^{free} (z)\,,
\end{align}
and the two surviving propagators in Eqs.~(\ref{QCD1})-(\ref{QCD2}) are considered free. Here we use $ A_\mu(z)=-\frac{1}{2}\, F_{\mu\nu}(z)\, z^\nu $ where  the electromagnetic field strength tensor is written as $ F_{\mu\nu}(z)=-i(\varepsilon_\mu q_\nu-\varepsilon_\nu q_\mu)\,e^{iq.z} $.  This amounts to taking $\bar T_4^{\gamma} (\underline{\alpha}) = 0$ and $S_{\gamma} (\underline {\alpha}) = \delta(\alpha_{\bar q})\delta(\alpha_{q})$ as the light-cone distribution amplitude in the three particle distribution amplitudes (see Ref. \cite{Li:2020rcg}). It is recommended that the procedure below be followed to obtain contributions from long distances,
 \begin{align}
\label{edmn21}
S_{q, \alpha\beta}^{ab}(x) \rightarrow -\frac{1}{4} \Big[\bar{q}^a(x) \Gamma_i q^b(0)\Big]\Big(\Gamma_i\Big)_{\alpha\beta},
\end{align}
and the two surviving propagators in Eqs.~(\ref{QCD1})-(\ref{QCD2}) are treated as full propagators.  Here $\Gamma_i$ $=$ $\{\textbf{1}$, $\gamma_5$, $\gamma_\mu$, $i\gamma_5 \gamma_\mu$, $\sigma_{\mu\nu}/2\}$. As the steps in equation (\ref{edmn21}) are carried out, new expressions, like $\langle \gamma(q)\vel \bar{q}(x) \Gamma_i G_{\alpha\beta}q(0) \ver 0\rangle$ and $\langle \gamma(q)\vel \bar{q}(x) \Gamma_i q(0) \ver 0\rangle$, become requirements for continuing the calculation.  These parameters, which are expressed in terms of photon wave functions, are of particular significance for the determination of long-distance effects (for details see Ref.~\cite{Ball:2002ps}). A full description of the procedures used to include the perturbative and non-perturbative contributions to the calculations can be found in Refs.~\cite{Ozdem:2022eds, Ozdem:2022vip}. The expressions in Eqs.~(\ref{QCD1})-(\ref{edmn21}) are used to calculate the QCD representation of the correlation functions. By applying the Fourier transform, these terms are then shifted into the momentum space. By using the above-mentioned lengthy and complicated steps, the QCD representation of the magnetic dipole moments is obtained.

The QCD light-cone sum rules for magnetic dipole moments can be constructed in such a way that the correlation functions obtained at the hadronic and quark-gluonic levels must be matched. As a final step, we employ continuum subtraction and Borel transformation schemes, according to the standard procedure of the QCD light-cone sum rules method. The results obtained for the magnetic dipole moments are illustrated in the following:
\begin{align}
\label{edmn15}
\mu^S_{B_Q}  &= \frac{e^{\frac{m^2_{ B_Q}}{\rm{M^2}}}}{\lambda^2_{ B_Q}\, m_{ B_Q} }\, \rho_1 (\rm{M^2},\rm{s_0}),~~~~~~
%%%%%%%%%%%%%%%%%%%%%%%%%%%%%%%%%%%%%
\mu^A_{B_Q}  =\frac{e^{\frac{m^2_{ B_Q}}{\rm{M^2}}}}{\lambda^2_{ B_Q}\, m_{ B_Q} }\, \rho_2 (\rm{M^2},\rm{s_0}).
%%%%%%%%%%%%%%%%%%%%%%%%%%%%%%%%%%%%%
\end{align}
The explicit forms of $\rho_i(\rm{M^2},\rm{s_0})$ expressions are as follows:
\begin{align}
\label{sumrules}
 \rho_1 (\rm{M^2},\rm{s_0}) &=\frac{m_Q^2}{32 \pi^4} \Big[3 m_Q^2 \Big ( (e_{q_1}+e_{q_2}+2e_{q_{12}}) (m_Q^2 I[-4,2]-I[-3,2])\Big) \Big]
 %%%%%%%%%%%%%%%%%%%%%%%%%%%%%%%%%%%%%%%%%%%%%%%%%
 \nonumber\\
 &- \frac{1}{64 m_Q^2 \pi^2 }\Big[ \Big(e_ {q_ 1} m_ {q_ 2} \langle \bar q_ 1 q_ 1 \rangle + 
     e_ {q_ 2} m_ {q_ 1} \langle \bar q_ 2 q_ 2 \rangle + 
     2 e_ {q_ {12}} m_ {q_ {12}} \langle \bar q_ {12} q_ {12} \rangle \Big) \Big ( \big (m_Q^4 I[-2, 0] - I[0, 0]\big) I_ 3[\mathcal S] \nonumber\\
     &+ 
    4 \chi \big (m_Q^6 I[-3, 1] + I[0, 1]\big) \varphi_ {\gamma}[
       u_ 0]\Big) -  f_ {3\gamma} \Big(e_ {q_ 1} + 2 e_ {q_ {12}} + 
    e_ {q_ 2}\Big)  \Big(m_Q^6 I[-3, 1] + 
    I[0, 1]\Big) \Big(8 I_ 2[\mathcal V] + \psi^a[u_ 0]\Big)
 \Big],\\
 %  \end{align}
%\begin{align}
  \rho_2 (\rm{M^2},\rm{s_0}) &=-\frac{3 e_Q m_Q^2}{32 \pi^4} \Big[m_Q^4 I[-4, 2] - 2 m_Q^2 I[-3, 2]+ I[-2, 2]\Big]
 %%%%%%%%%%%%%%%%%%%%%%%%%%%%%%%%%%%%%%%%%%%%%%%%%
 \nonumber\\
 &+\frac{1}{64 m_Q^2 \pi^2 }\Big[2 \Big (e_ {q_ 1} m_ {q_ 2}\langle \bar q_ 1 q_ 1 \rangle + 
    e_ {q_ 2} m_ {q_ 1}\langle \bar q_ 2 q_ 2 \rangle + 
    2 e_ {q_ {12}} m_ {q_ {12}}\langle \bar q_ {12} q_ {12} \rangle\Big) (m_Q^4 I[-2, 0] - I[0, 0]) I_ 3[\mathcal S] \nonumber\\
    &+ 
 f_ {3\gamma} m_Q^4 \Big (e_ {q_ 1} + 2 e_ {q_ {12}} + 
    e_ {q_ 2}\Big)  (m_Q^2 I[-3, 1] + I[-2, 1]) I_ 2[\mathcal V]\Big],
 \label{sumrules2}
\end{align}
where the functions $I_i[\mathcal{F}]$ and $I[n,m]$ are defined as follows:
\begin{align}
I[n,m]&= \int_{\mathcal{M}^2}^{\mathrm{s_0}} ds \int_{\mathcal{M}^2}^s dw \,w^n~(s-w)^m\, e^{-s/\mathrm{M^2}},\nonumber\\
%   \end{align}
% \begin{align}
 I_2[\mathcal{F}]&=\int D_{\alpha_i} \int_0^1 dv~ \mathcal{F}(\alpha_{\bar q},\alpha_q,\alpha_g) \delta'(\alpha_ q +\bar v \alpha_g-u_0),\nonumber\\
  I_3[\mathcal{F}]&=\int D_{\alpha_i} \int_0^1 dv~ \mathcal{F}(\alpha_{\bar q},\alpha_q,\alpha_g) \delta'(\alpha_{\bar q}+ v \alpha_g-u_0),%\nonumber%\\
%    I_3[\mathcal{F}]&=\int D_{\alpha_i} \int_0^1 dv~ \mathcal{F}(\alpha_{\bar q},\alpha_q,\alpha_g) \delta(\alpha_ q +\bar v \alpha_g-u_0),\nonumber\\
%   I_4[\mathcal{F}]&=\int D_{\alpha_i} \int_0^1 dv~ \mathcal{F}(\alpha_{\bar q},\alpha_q,\alpha_g) \delta(\alpha_{\bar q}+ v \alpha_g-u_0),\nonumber\\
 %\end{align}
 %\begin{align}
%   I_2[\mathcal{F}]&=\int_0^1 du~ \mathcal{F}(u)\delta'(u-u_0),%\nonumber\\
 %I_6[\mathcal{F}]&=\int_0^1 du~ \mathcal{F}(u),
 \end{align}
 where $\mathcal M = m_Q^2$ for the $\Sigma_Q$ and $\Lambda_Q$ baryons; $\mathcal M = (m_Q+m_s)^2$ for the $\Xi_Q$ and $\Xi^\prime_Q$ baryons; $\mathcal M = (m_Q+2m_s)^2$ for the $\Omega_Q$ baryons,  and $\mathcal{F}$ being the relevant photon DAs. Note that the terms $e_{q_{12}}$, $m_{q_{12}}$, and $\langle \bar q_{12} q_{12} \rangle $ exist only if $q_1=q_2$.

\subsection{Formalism of the spin-$\frac{3}{2}$ singly-heavy baryons}

This subsection of the paper will undertake a magnetic dipole, electric quadrupole, and magnetic octupole moments analysis of singly-heavy baryons with a spin-parity quantum number of $\rm{J^P} = \frac{3}{2}^+$. As the methodology was elucidated in the preceding subsection, this section will not provide a comprehensive overview; instead, it will adopt a similar approach to that previously described. The initial stage of the process is to define the relevant correlation function, which is expressed as follows:
 \begin{align} \label{edmn001}
\Pi_{\mu\nu}(p,q)&=i\int d^4x e^{ip \cdot x} \langle0|T\left\{\rm{J_\mu}(x)\bar{\rm{J_\nu}}(0)\right\}|0\rangle _\gamma \, , %\\
\end{align}
where $\mathrm{J_\mu}(x)$ is the interpolating currents of the singly-heavy baryons with $\rm{J^P} = \frac{3}{2}^+$ and it is given as,
\begin{align}
\rm{J}_{\mu}(x)& =   \varepsilon^{abc} \big[q_1^{a^T}(x) C\gamma_\mu q_2^b(x)\big] 
  Q^c(x) \, , \label{curpcs3}
%%%%%%%%%%%%%%%%%%%%%%%%%%%%%%%%%%%%%%%%%%%%%%%%%%%%%%%%%%%%%%%%%%%%%%%%%%%%%%%%%%%
\end{align}
where $a$, $b$, and $c$ being color indices; $Q= c$ and $b$-quark;  and the $C$ is the charge conjugation operator. The quark content of the spin-$\frac{3}{2}$ singly-heavy baryons is presented in Table \ref{quarkcon1}.
\begin{table}[htp]
	\addtolength{\tabcolsep}{10pt}
		\begin{center}
		\caption{The quark content of the spin-$\frac{3}{2}$ singly-heavy baryons.}
	\label{quarkcon1}
\begin{tabular}{lccccccccc}
	   \hline\hline
	  % \\
  Quarks& $\Sigma_{c(b)}^{*++(+)}$&$\Sigma_{c(b)}^{*+(0)}$& $\Sigma_{c(b)}^{*0(-)}$&
  $\Xi_{ c(b)}^{* +(0)}$& $\Xi_{ c(b)}^{* 0(-)}$& $\Omega_{c(b)}^{*0(-)}$\\
  %\\
\hline\hline
%\\
$q_1$&  u & u & d & u & d & s\\
%\\
$q_2$&  u & d & d & s & s & s\\
%\\
	   \hline\hline
\end{tabular}
\end{center}
\end{table}

The correlation function is written in terms of hadron parameters as follows:
\begin{eqnarray}\label{edmn002}
\Pi^{Had}_{\mu\nu}(p,q)&=&\frac{\langle0\mid J_{\mu}(x)\mid
B_Q^*(p_2)\rangle}{[p_2^{2}-m_{B_Q^*}^{2}]}\langle B_Q^*(p_2)\mid
B_Q^*(p_1)\rangle_\gamma\frac{\langle B_Q^*(p_1)\mid
\bar{J}_{\nu}(0)\mid 0\rangle}{[p_1^{2}-m_{B_Q^*}^{2}]},
\end{eqnarray}
where $p_1 = p+q$, $p_2=p$. As can be seen, the aforementioned equation contains matrix elements that are necessary for further analysis. The explicit forms of these matrix elements are as follows:%~\cite{Weber:1978dh,Nozawa:1990gt,Pascalutsa:2006up,Ramalho:2009vc}:

\begin{eqnarray}
\label{lambdabey}
\langle0\mid J_{\mu}(x)\mid B_Q^*(p_2,s)\rangle &=&\lambda_{B_Q^*}u_{\mu}(p_2,s),\\
%\nonumber\\
%%%%%%%%%%%%%%%%%%%%%%%%%%%%%%%%%%%%%%%%%%%%%%%%%%%%%%%%%%%%%%%%%%%%
\langle {B_Q^*}(p_1,s)\mid
\bar{J}_{\nu}(0)\mid 0\rangle &=& \lambda_{{B_Q^*}}\bar u_{\nu}(p_1,s),\\
%\nonumber\\
%%%%%%%%%%%%%%%%%%%%%%%%%%%%%%%%%%%%%%%%
\langle B_Q^*(p_2)\mid B_Q^*(p_1)\rangle_\gamma &=&-e \,\bar
u_{\mu}(p_2)\left\{F_{1}(q^2)g_{\mu\nu}\eslash-
\frac{1}{2m_{B_Q^*}}\left
[F_{2}(q^2)g_{\mu\nu} \eslash\qslash+F_{4}(q^2)\frac{q_{\mu}q_{\nu} \eslash\qslash}{(2m_{B_Q^*})^2}\right]
\right.\nonumber\\&+&\left.
\frac{F_{3}(q^2)}{(2m_{B_Q^*})^2}q_{\mu}q_{\nu}\eslash\right\} u_{\nu}(p_1),\label{matelpar}
\end{eqnarray}
where $\lambda_{B_Q^*}$  is residue of the $B_Q^*$ states; $u_{\mu}(p_2,s)$ and $ \bar u_{\nu}(p_1,s)$ are the spinors of the $B_Q^*$ states; %$\varepsilon$ is the photon's polarization vector, 
and $F_i (q^2)$ are transition form factors.  Using the equations (\ref{edmn002})-(\ref{matelpar}) and after performing the necessary mathematical manipulations, the expressions of the electromagnetic multipole moments of $B_Q^*$ baryons concerning the hadronic parameters are as follows:
\begin{eqnarray}
\label{final phenpart}
\Pi^{Had}_{\mu\nu}(p,q)&=&-\frac{\lambda_{_{B_Q^*}}^{2}}{[(p+q)^{2}-m_{_{B_Q^*}}^{2}][p^{2}-m_{_{B_Q^*}}^{2}]}
\bigg[2 \,\varepsilon_{\mu} q_{\nu}\qslash \,F_{1}(q^2) 
+2\, q_{\mu} q_{\nu}\eslash \,F_{2}(q^2)- \frac{1}{2 m_{B^*}} q_{\mu} q_{\nu}\eslash\qslash  F_3(q^2) 
\nonumber\\
&&
- \frac{1}{4 m^2_{B^*}} (\varepsilon. p) q_{\mu} q_{\nu}\qslash F_4(q^2)
+
\mathrm{other~independent~structures} \bigg].
\end{eqnarray}
It is preferable to utilize the $\varepsilon_{\mu} q_{\nu}\qslash$,  $q_{\mu} q_{\nu}\eslash$, $q_{\mu} q_{\nu}\eslash\qslash$, and $(\varepsilon. p) q_{\mu} q_{\nu}\qslash$ structures, as they effectively exclude the possibility of spin-1/2 baryons contributing to the structures under consideration.

It may be more meaningful to express the form factors given in Eq.~(\ref{final phenpart}) regarding the magnetic dipole, $G_{M}(q^2)$, electric quadrupole, $G_{Q}(q^2)$, and magnetic octupole, $G_{O}(q^2)$, form factors, given that these are experimentally accessible quantities. The expression for these are provided below~\cite{Weber:1978dh,Nozawa:1990gt,Pascalutsa:2006up,Ramalho:2009vc}: 
\begin{eqnarray}
G_{M}(q^2) &=& \left[ F_1(q^2) + F_2(q^2)\right] ( 1+ \frac{4}{5}
\tau ) -\frac{2}{5} \left[ F_3(q^2)  +
F_4(q^2)\right] \tau \left( 1 + \tau \right), \nonumber\\
G_{Q}(q^2) &=& \left[ F_1(q^2) -\tau F_2(q^2) \right]  -
\frac{1}{2}\left[ F_3(q^2) -\tau F_4(q^2)
\right] \left( 1+ \tau \right),  \nonumber \\
 G_{O}(q^2) &=&
\left[ F_1(q^2) + F_2(q^2)\right] -\frac{1}{2} \left[ F_3(q^2)  +
F_4(q^2)\right] \left( 1 + \tau \right),
\end{eqnarray}
  where $\tau
= -\frac{q^2}{4m^2_{B_Q^*}}$.  At $q^2=0$, the electromagnetic multipole form factors  regarding the $F_i(0)$ form factors is given by
\begin{eqnarray}\label{mqo1}
G_{M}(0)&=&F_{1}(0)+F_{2}(0),\nonumber\\
G_{Q}(0)&=&F_{1}(0)-\frac{1}{2}F_{3}(0),\nonumber\\
G_{O}(0)&=&F_{1}(0)+F_{2}(0)-\frac{1}{2}[F_{3}(0)+F_{4}(0)].
\end{eqnarray}
 
 Given the focus of our study on the analysis of magnetic dipole, electric quadrupole, and magnetic octupole moments, it is imperative that these electromagnetic multipole moments are expressed following the previously mentioned form factors.  
The magnetic dipole ($\mu_{B^*_Q}$), electric quadrupole ($Q_{B^*_Q}$) and magnetic octupole ($O_{B^*_Q}$) moments can be derived from the aforementioned term through the following process: 

 \begin{eqnarray}\label{mqo2}
\mu_{B^*_Q}&=&\frac{e}{2m_{B^*_Q}}G_{M}(0),~~~~~~%\nonumber\\
Q_{B^*_Q}=\frac{e}{m_{B^*_Q}^2}G_{Q}(0), ~~~~~~% \nonumber\\
O_{B^*_Q}=\frac{e}{2m_{B^*_Q}^3}G_{O}(0).
\end{eqnarray}

Having completed the analysis of the correlation function for hadronic parameters, we may now proceed with the analysis of the corresponding QCD parameters. The result of the correlation function for the relevant QCD parameters is as follows:
\begin{align}
\label{QCD3}
\Pi_{\mu\nu}^{\rm{QCD}}(p,q)&= i\varepsilon^{abc} \varepsilon^{a^{\prime}b^{\prime}c^{\prime}}\, \int d^4x \, e^{ip\cdot x} 
 \langle 0\mid \Big\{ \mbox{Tr}\Big[\gamma_{\mu} S_{q_2}^{bb^\prime}(x) \gamma_{\nu}  
  \widetilde S_{q_1}^{aa^\prime}(x)\Big]S_{Q}^{cc^\prime}(x)  \nonumber\\
&-
\mbox{Tr}\Big[\gamma_{\mu} S_{q_2 q_1}^{ba^\prime}(x) \gamma_{\nu}  
  \widetilde S_{q_1 q_2}^{ab^\prime}(x)\Big] S_{Q}^{cc^\prime}(x)
  \Big\} 
\mid 0 \rangle _\gamma \,. 
\end{align}

We have now derived analytical expressions for the correlation function in terms of both QCD and hadron parameters. The subsequent stage of the process will be to ensure that the results obtained in the two distinct regions are aligned. Upon completion of this process, the magnetic dipole moments of singly-heavy baryons with $\rm{J^P}=\frac{3}{2}^+$ are found from
\begin{align}
\label{spin32heavy}
 \mu_{B_Q^*}  &= \frac{e^{\frac{m^2_{B_Q^*}}{\mathrm{M^2}}}}{\lambda^2_{B_Q^*}}\, \rho_3 (\mathrm{M^2},\mathrm{s_0}), ~~~~
 \mathcal{Q}_{B_Q^*}  = \frac{e^{\frac{m^2_{B_Q^*}}{\mathrm{M^2}}}}{\lambda^2_{B_Q^*}}\, \rho_4 (\mathrm{M^2},\mathrm{s_0}), 
 ~~~~
 \mathcal{O}_{B_Q^*}  = \frac{e^{\frac{m^2_{B_Q^*}}{\mathrm{M^2}}}}{\lambda^2_{B_Q^*}}\, \rho_5 (\mathrm{M^2},\mathrm{s_0}),
\end{align}
where 
\begin{align}
\rho_3 (\mathrm{M^2},\mathrm{s_0})&=\frac{1}{2}\Big[ F_1(\mathrm{M^2},\mathrm{s_0}) +F_2(\mathrm{M^2},\mathrm{s_0})\Big], \\
       %%%%%%%%%%%%%%%%%%%%%%%%%%%%%%%%%
\rho_4 (\mathrm{M^2},\mathrm{s_0})&=\Big[ \frac{1}{2}F_1(\mathrm{M^2},\mathrm{s_0}) + m_{B_Q^*}F_3(\mathrm{M^2},\mathrm{s_0})\Big], \\
       %%%%%%%%%%%%%%%%%%%%%
\rho_5 (\mathrm{M^2},\mathrm{s_0})&=\Big[ \frac{1}{2}F_1(\mathrm{M^2},\mathrm{s_0}) +\frac{1}{2}F_2(\mathrm{M^2},\mathrm{s_0})+ m_{B_Q^*}F_3(\mathrm{M^2},\mathrm{s_0})+ 2  m^2_{B_Q^*}F_4(\mathrm{M^2},\mathrm{s_0})\Big]. 
      \end{align}

The explicit forms of the $F_1(\mathrm{M^2},\mathrm{s_0})$, $F_2(\mathrm{M^2},\mathrm{s_0})$, $F_3(\mathrm{M^2},\mathrm{s_0})$ and $F_4(\mathrm{M^2},\mathrm{s_0})$ form factors are given as;
\begin{align}
\label{F1son}
 F_1(\mathrm{M^2},\mathrm{s_0})&= \frac{m_Q^2}{256 \pi^4} \Big[3m_Q^2 (e_ {q_ 1} + e_ {q_ 2}+2e_ {q_ {12}})  \Big (m_Q^4 \big(I[-5, 2] + 2 I[-4, 1]\big) + 
   2 m_Q^2 \big(I[-4, 2] + 2 I[-3, 1]\big) - 3 I[-3, 2] \nonumber\\
   &+ 2 I[-2, 1]\Big) +
  4 e_Q\Big (m_Q^6 I[-5, 2] - 3 m_Q^4 I[-4, 2] + 3 m_Q^2 I[-3, 2] - 
   I[-2, 2]\Big)\Big]
        %%%%%%%%%%%%%%%%%%%%%%%%%%%%%%%%%%%%%%
        \nonumber\\
       % \end{align}
       % \begin{align}
          &-\frac{1}{64 m_Q^2 \pi^2}
          \Big(\langle \bar q_1 q_1 \rangle  e_{q_1} m_{q_2} + \langle \bar q_2 q_2 \rangle e_{q_2} m_{q_1} + 2 e_{q_{12}} m_{q_{12}} \langle \bar q_{12} q_{12} \rangle\Big)
  \Big[
    \big (m_Q^4 I[-2, 0] - I[0, 0]\big) I_ 3[\mathcal T_ 1] \nonumber\\
 &- 
 2 I[0, 0] I_ 6[h_ {\gamma}]  + 
 4 \chi \big (m_Q^6 I[-3, 1] + I[0, 1]) \varphi_{\gamma}[u_ 0]
  \Big]
  %%%%%%%%%%%%%%%%%%%%%%%%%%%%%%%%%%%5
  \nonumber\\
 % \end{align}
 %       \begin{align}
  &-\frac{f_ {3\gamma}}{128 m_Q^2 \pi^2} \Big( e_{q_1}+2 e_{q_{12}}+ e_{q_2}\Big )\Big[m_Q^6 \Big (I_ 2[\mathcal A] - I_ 2[\mathcal V] + I_ 3[\mathcal A] - 
   I_ 3[\mathcal V]\Big) I[-3, 1]
   -8 m_Q^6 \big (m_Q^2 I[-4, 1] \nonumber\\
   & + 
   I[-3, 1]\big) I_ 6[\psi_ {\gamma}^{\nu}]
   +
   2 m_Q^4 \big(I_3[\mathcal A] - I_3[\mathcal V]\big) I[-2, 1]
   + \big(m_Q^8 I[-4, 1] + m_Q^6 I[-3, 1] + 2 I[0, 1]\big)I_5[\psi^a]
    \nonumber\\
   &+
   2 \big(3 m_Q^8 I[-4, 1]+ m_Q^6 I[-3, 1] - 2 I[0, 1]\big) \psi^a[u_0]
   -4 m_Q^6 \big(m_Q^2 I[-4, 1] + I[-3, 1]\big) \psi_ {\gamma}^{\nu}[u_0]
  \Big],\\
  \nonumber\\
  %\nonumber\\
%  \end{align}
%\begin{align}
 F_2(\mathrm{M^2},\mathrm{s_0})&=
 \frac{e_Q}{32 \pi^4} \Big[ m_Q^8 \big (I[-5, 2] + I[-4, 1]\big) - 
 3 m_Q^6 \big (I[-4, 2] - I[-3, 1]\big) + 
 3 m_Q^4 \big (I[-3, 2] + I[-2, 1]\big)\nonumber\\
 &+ 
 m_Q^2 \big (-I[-2, 2] + I[-1, 1]\big) + 8 I[0, 1]\Big]
        %%%%%%%%%%%%%%%%%%%%%%%%%%%%%%%%%%%%%%
        \nonumber\\
  &+\frac{1}{32 m_Q^2 \pi^2}
  \Big(\langle \bar q_1 q_1 \rangle  e_{q_1} m_{q_2} + \langle \bar q_2 q_2 \rangle e_{q_2} m_{q_1} + 2 e_{q_{12}} m_{q_{12}} \langle \bar q_{12} q_{12} \rangle\Big)
  \Big[  \big(m_Q^4 I[-2, 0] - I[0, 0]\big) I_3[\mathcal T_1]
  \Big],\nonumber\\
  &
  +\frac{f_ {3\gamma}}{64 m_Q^2 \pi^2} \Big( e_{q_1}+2 e_{q_{12}}+ e_{q_2}\Big )\Big[
   8 m_Q^8 I_ 6[\psi_ {\gamma}^{\nu}] I[-4, 1] + 
 m_Q^6 \Big (3 I_ 3[\mathcal A]-2 I_ 2[\mathcal V]  + 
    4 I_ 3[\mathcal V] + 8 I_ 6[\psi_{\gamma}^{\nu}]\Big) I[-3, 1]\nonumber\\
    &+ 
 2 m_Q^4 \Big ( I_ 3[\mathcal A]-I_ 2[\mathcal V] + 
    2 I_ 3[\mathcal V]\Big) I[-2, 1] + I_ 3[\mathcal A] I[0, 1]
  \Big], \label{F2son}\\ 
  \nonumber\\
%\end{align}
%\begin{align}
  F_3(\mathrm{M^2},\mathrm{s_0})&=-
 \frac{e_Q}{16 \,m_Q\, \pi^4} \Big[ m_Q^2 \big(m_Q^6 I[-4, 1] + 3 m_Q^4 I[-3, 1] + 3 m_Q^2 I[-2, 1] + 
  I[-1, 1] \big) + 8 I[0, 1]\Big]
        %%%%%%%%%%%%%%%%%%%%%%%%%%%%%%%%%%%%%%
        \nonumber\\
  &
  +\frac{f_ {3\gamma}\,m_Q}{32 \pi^2} \Big( e_{q_1}+2 e_{q_{12}}+ e_{q_2}\Big )\Big[  \big( m_Q^2 I[2, 0] -I[-1, 0] \big ) I_3[\mathcal V]
  \Big], \label{F3son} \\ 
  \nonumber\\
  %%%%%%%%%%%%%%%%%%%%%%%%%%%%%%%5
  %%%%%%%%%%%%%%%%%%%%%%%%%%%%%%%%%
   F_4(\mathrm{M^2},\mathrm{s_0})&=-
 \frac{e_Q}{16 \,m_Q^2\, \pi^4} \Big[m_Q^{4} \big ( m_Q^6 I[-5, 1] + 3 m_Q^4 I[-4, 1] + 3 m_Q^2 I[-3, 1] + 
  I[-2, 1]  \big)+ 8 I[0, 1]\Big]
        %%%%%%%%%%%%%%%%%%%%%%%%%%%%%%%%%%%%%%
        \nonumber\\
  &
  -\frac{f_ {3\gamma}}{32\,m_Q^2 \pi^2} \Big( e_{q_1}+2 e_{q_{12}}+ e_{q_2}\Big )\Big[  
  I_3[\mathcal A] \,I[0, 0] + 
 2  \big( m_Q^6 I[3, 0]-m_Q^4 I[-2, 0] \big) I_3[\mathcal V] + 
 4 m_Q^8 I_6[\psi_{\gamma}^{\nu}] I[4, 0]
  \Big], \label{F4son}
\end{align}
with
\begin{align}
I[n,m]&= \int_{\mathcal{M}^2}^{\mathrm{s_0}} ds \int_{\mathcal{M}^2}^s dw \,w^n~(s-w)^m\, e^{-s/\mathrm{M^2}},\nonumber\\
%   \end{align}
% \begin{align}
 I_2[\mathcal{F}]&=\int D_{\alpha_i} \int_0^1 dv~ \mathcal{F}(\alpha_{\bar q},\alpha_q,\alpha_g) \delta'(\alpha_ q +\bar v \alpha_g-u_0),\nonumber\\
   I_3[\mathcal{F}]&=\int D_{\alpha_i} \int_0^1 dv~ \mathcal{F}(\alpha_{\bar q},\alpha_q,\alpha_g) \delta'(\alpha_{\bar q}+ v \alpha_g-u_0),\nonumber\\
 %\end{align}
 %\begin{align}
   I_5[\mathcal{F}]&=\int_0^1 du~ \mathcal{F}(u)\delta'(u-u_0),\nonumber\\
 I_6[\mathcal{F}]&=\int_0^1 du~ \mathcal{F}(u),
 \end{align}
 where $\mathcal M = m_Q^2$ for the $\Sigma_Q^*$  baryons; $\mathcal M = (m_Q+m_s)^2$ for the $\Xi_Q^*$ baryons; $\mathcal M = (m_Q+2m_s)^2$ for the $\Omega_Q^*$ baryons,  and $\mathcal{F}$ being the relevant photon DAs. Note that the terms $e_{q_{12}}= e_{q_1 q_2}$, $m_{q_{12}}= m_{q_1 q_2}$, and $\langle \bar q_{12} q_{12} \rangle$ exist only if $q_1=q_2$.

\end{widetext}

\section{Numerical Analysis}\label{numerical} 

In this section, we will perform numerical analyses of the expressions related to the magnetic dipole moments for singly-heavy baryons. To do so, we require a set of input parameters, which are taken as $m_s= 93.5 \pm 0.8$ MeV, $m_c= 1.273 \pm 0.0046$ GeV, $m_b= 4.183 \pm 0.007$ GeV~\cite{ParticleDataGroup:2024cfk}, $\chi= -2.85 \pm 0.5 $ GeV$^{-2}$ \cite{Rohrwild:2007yt}, and $\langle \bar ss\rangle = 0.8 \langle \bar qq\rangle$ with  $\langle \bar qq\rangle = (-0.24 \pm 0.01)^3 $ GeV$^3$ \cite{Ioffe:2005ym}. In numerical analysis, we set $m_u =m_d = 0$ and $m^2_s = 0$, but consider terms proportional to $m_s$.    To perform a numerical analysis, it is necessary to have the mass and residue values of the singly-heavy baryons. The values for the mass are taken from the Particle Data Group (PDG) \cite{ParticleDataGroup:2024cfk}, while the residue values are taken from the Refs.~\cite{Wang:2009cr, Wang:2020mxk,Wang:2010vn}. From sum rules in Eqs. (\ref{sumrules})-(\ref{sumrules2}) and Eqs.~(\ref{F1son})-(\ref{F4son}), it follows that the photon DAs are also required. Their explicit expressions are borrowed from Ref.~\cite{Ball:2002ps}. In addition to the parameters mentioned above, two extra parameters result from the systematics of the method itself: The Borel mass parameters $\rm{M^2}$ and continuum threshold parameters $\rm{s_0}$. Technically speaking, an interval should be chosen in which the variation of the obtained magnetic dipole moments concerning these extra parameters is relatively small and analyzed accordingly. Considering these conditions, the intervals determined for these parameters for the magnetic dipole moments of singly-heavy baryons, where the variation of the magnetic dipole moment results concerning these parameters is minimal, are given in Tables \ref{parameter} and \ref{parameter2}.

All requisite input parameters for the numerical analysis of the magnetic dipole moments of singly-heavy baryons have been identified and defined. The magnetic dipole moments obtained from the numerical analysis, taking into account these parameters and their associated uncertainties, are presented in Tables \ref{parameter} and \ref{parameter2}. 

\begin{figure}[htb!]
%\centering
\subfloat[]{\includegraphics[width=0.35\textwidth]{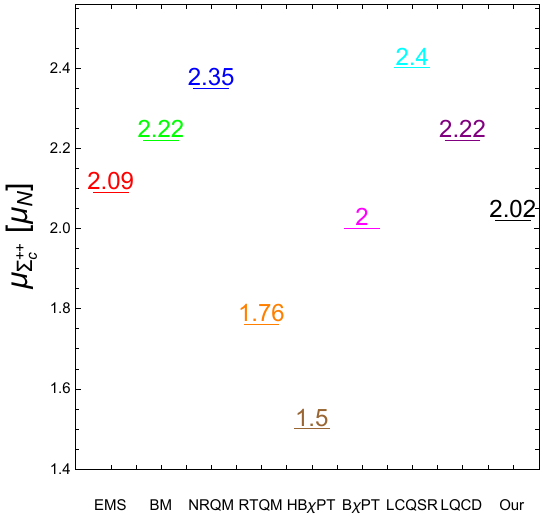}}~~~~~~~~~~
\subfloat[]{\includegraphics[width=0.35\textwidth]{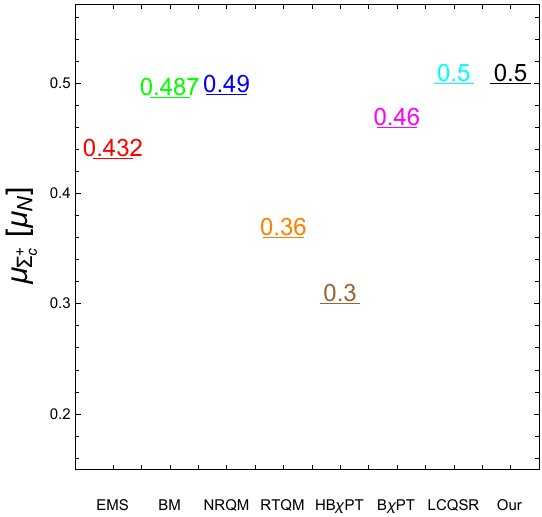}}\\
\subfloat[]{\includegraphics[width=0.35\textwidth]{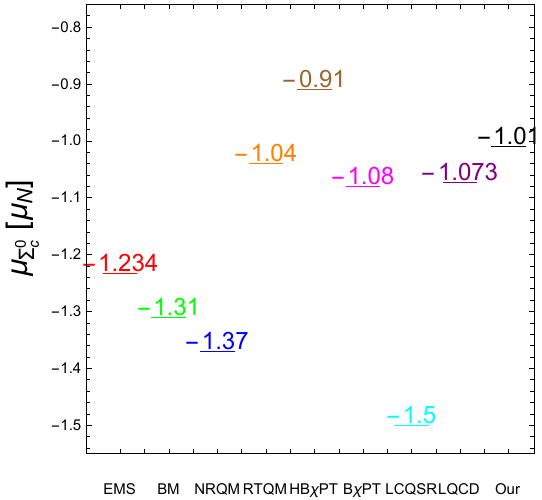}}~~~~~~~~~~
\subfloat[]{\includegraphics[width=0.35\textwidth]{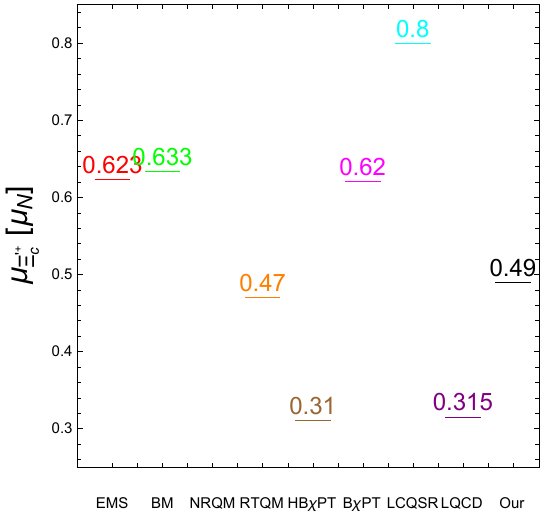}}\\
\subfloat[]{\includegraphics[width=0.35\textwidth]{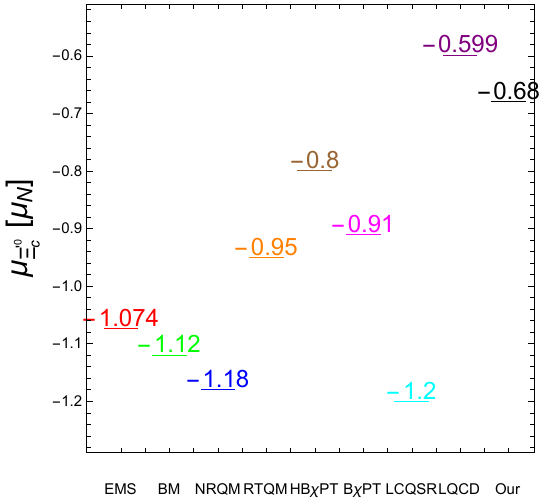}}~~~~~~~~~~
\subfloat[]{\includegraphics[width=0.35\textwidth]{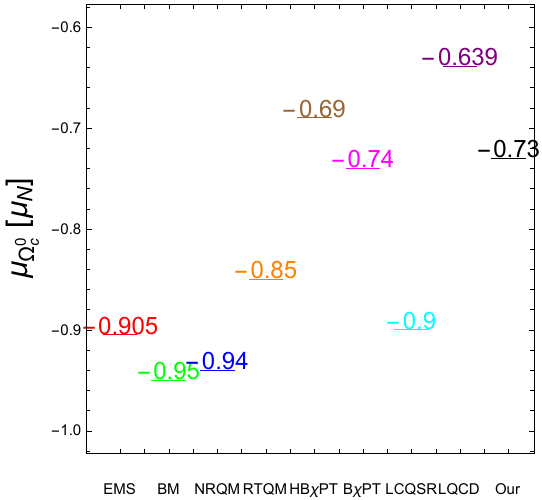}}
 \caption{Magnetic dipole moments of the $\rm{J^P} =\frac{1}{2}^+$ singly-charm sextet baryons obtained in different approaches.}
 \label{comp1}
  \end{figure}

\begin{figure}[htb!]
%\centering
\subfloat[]{\includegraphics[width=0.35\textwidth]{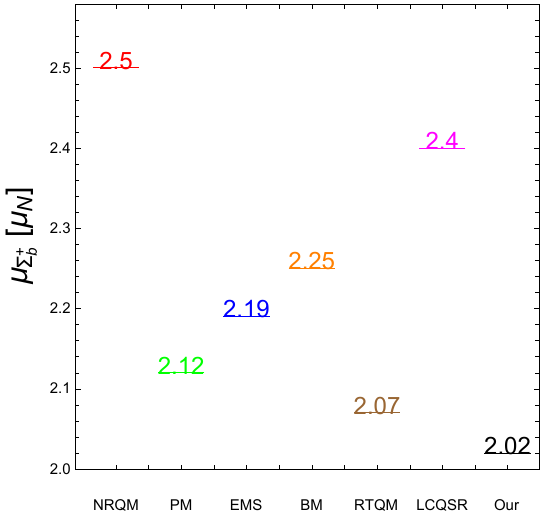}}~~~~~~~~~~
\subfloat[]{\includegraphics[width=0.35\textwidth]{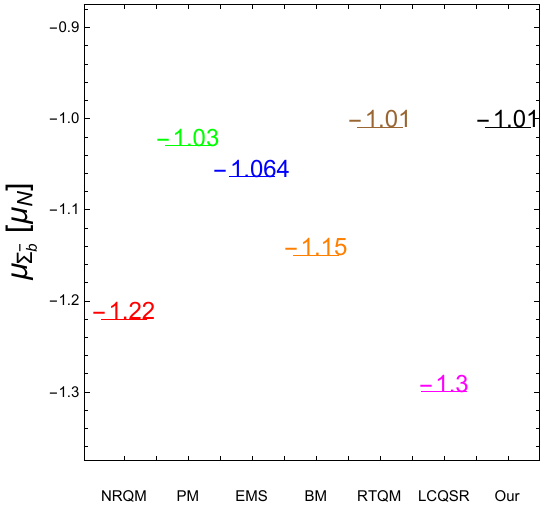}}\\
\subfloat[]{\includegraphics[width=0.35\textwidth]{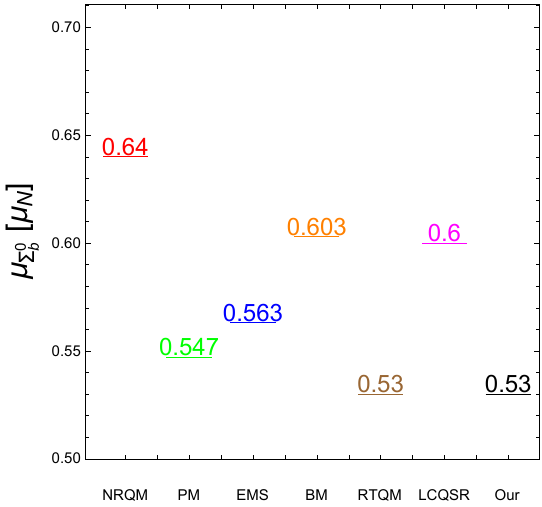}}~~~~~~~~~~
\subfloat[]{\includegraphics[width=0.35\textwidth]{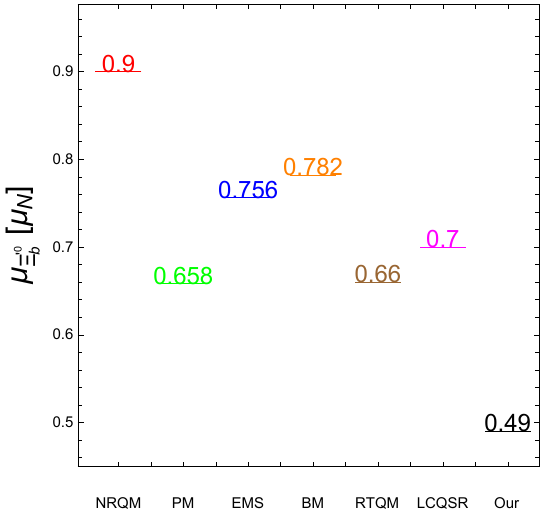}}\\
\subfloat[]{\includegraphics[width=0.35\textwidth]{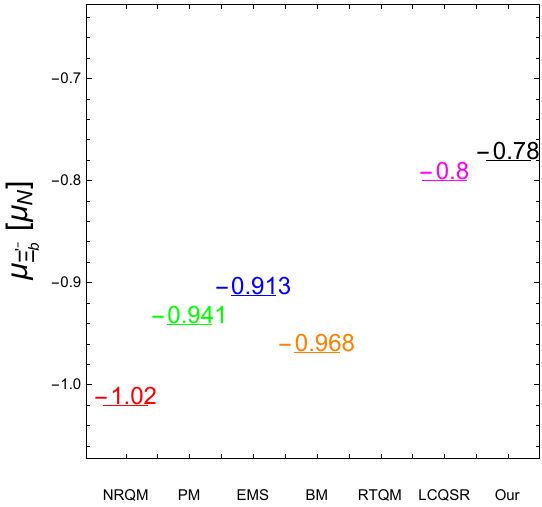}}~~~~~~~~~~
\subfloat[]{\includegraphics[width=0.35\textwidth]{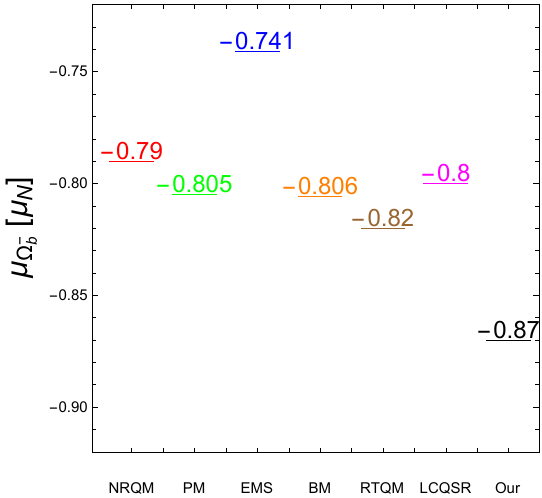}}
 \caption{Magnetic dipole moments of the $\rm{J^P} =\frac{1}{2}^+$ singly-bottom sextet baryons obtained in different approaches.}
 \label{comp2}
  \end{figure}

\begin{figure}[htb!]
%\centering
\subfloat[]{\includegraphics[width=0.35\textwidth]{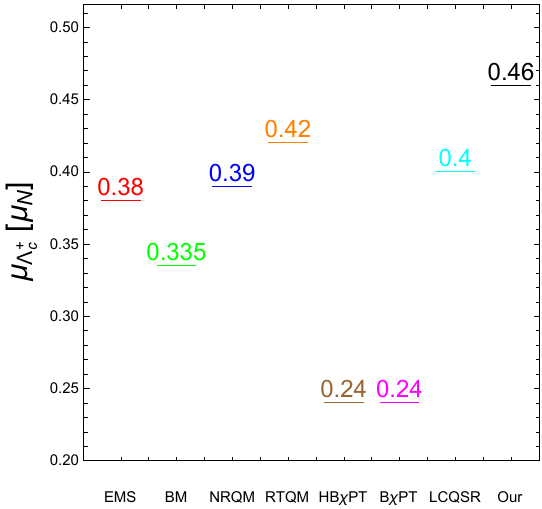}}~~~~~~~~~~
\subfloat[]{\includegraphics[width=0.35\textwidth]{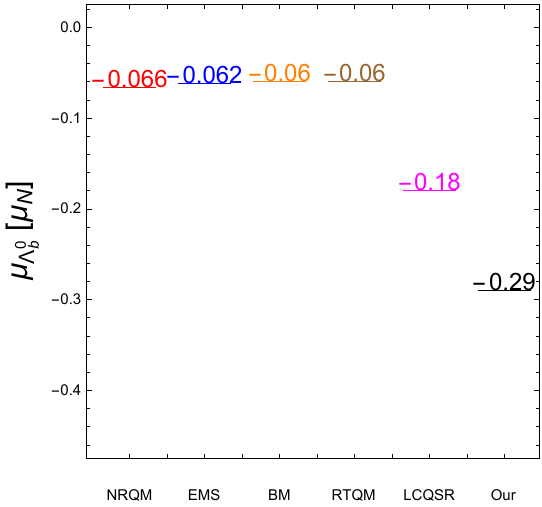}}\\
\subfloat[]{\includegraphics[width=0.35\textwidth]{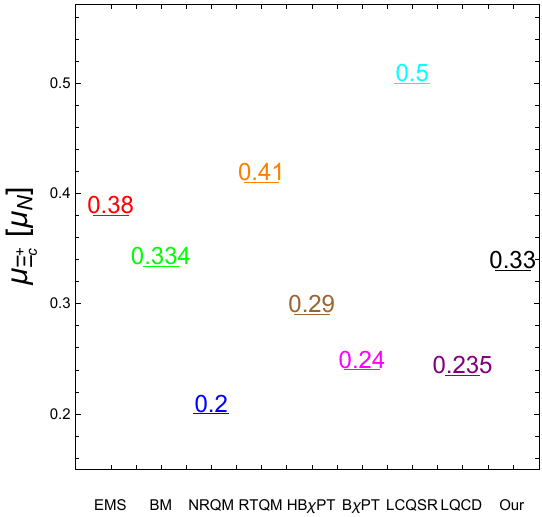}}~~~~~~~~~~
\subfloat[]{\includegraphics[width=0.35\textwidth]{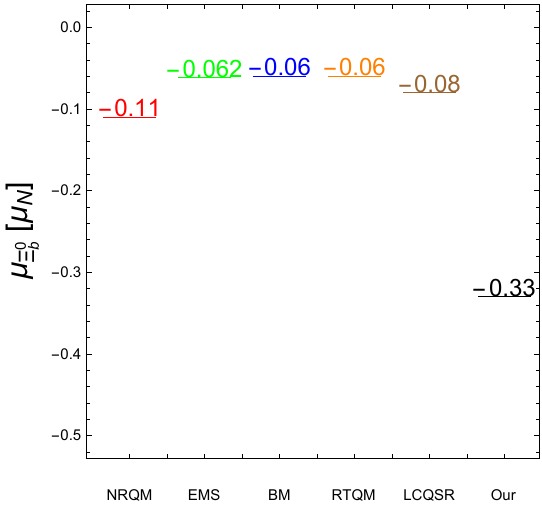}}\\
\subfloat[]{\includegraphics[width=0.35\textwidth]{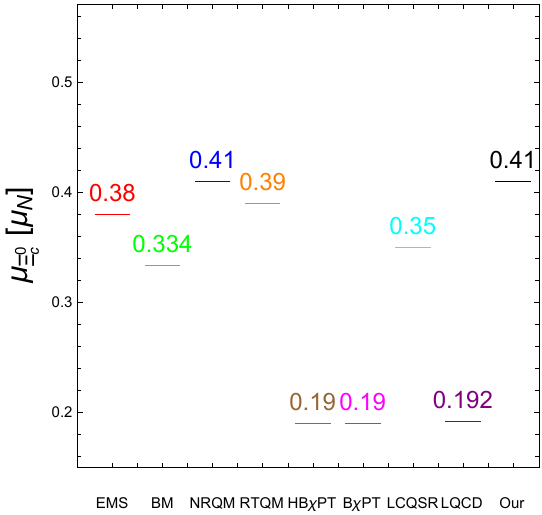}}~~~~~~~~~~
\subfloat[]{\includegraphics[width=0.35\textwidth]{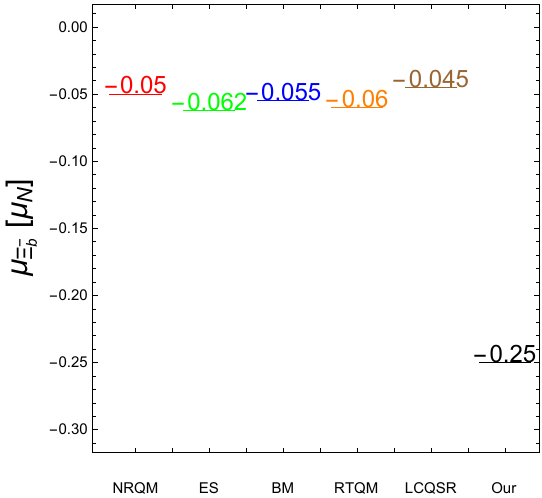}}
 \caption{Magnetic dipole moments of the $\rm{J^P} =\frac{1}{2}^+$ anti-triplet singly-charm and singly-bottom baryons obtained in different approaches.}
 \label{comp3}
  \end{figure}

\begin{figure}[htb!]
%\centering
\subfloat[]{\includegraphics[width=0.35\textwidth]{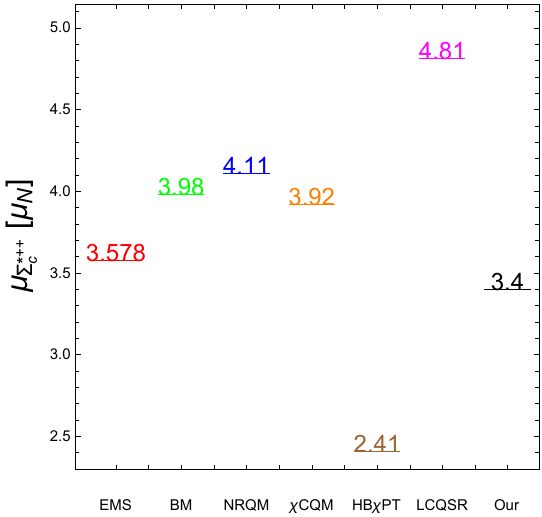}}~~~~~~~~~~
\subfloat[]{\includegraphics[width=0.35\textwidth]{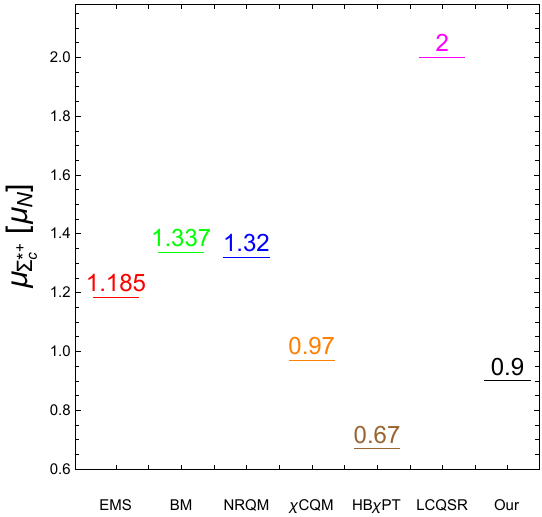}}\\
\subfloat[]{\includegraphics[width=0.35\textwidth]{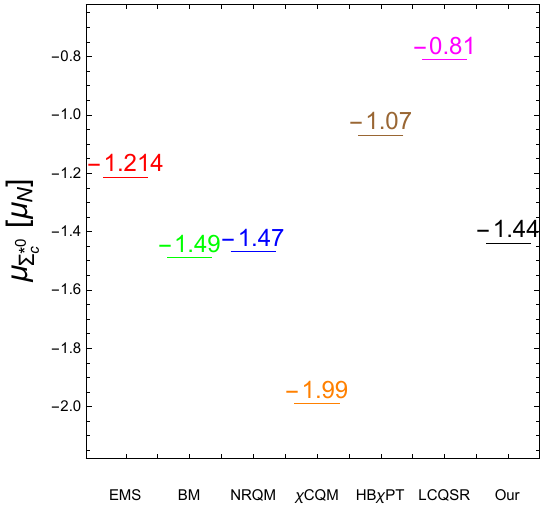}}~~~~~~~~~~
\subfloat[]{\includegraphics[width=0.35\textwidth]{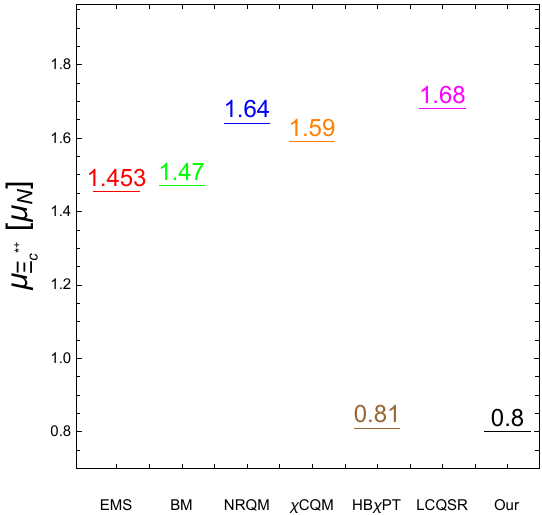}}\\
\subfloat[]{\includegraphics[width=0.35\textwidth]{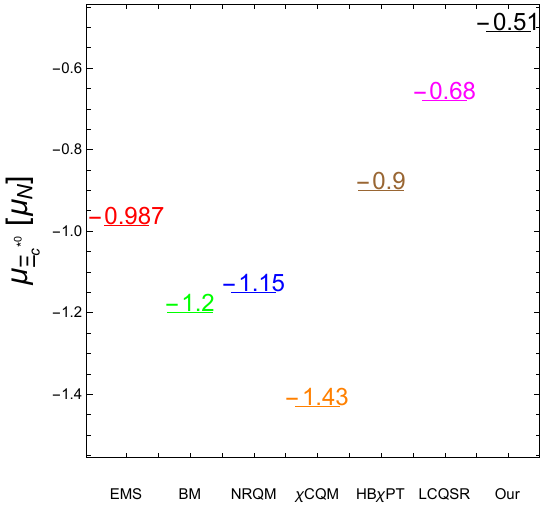}}~~~~~~~~~~
\subfloat[]{\includegraphics[width=0.35\textwidth]{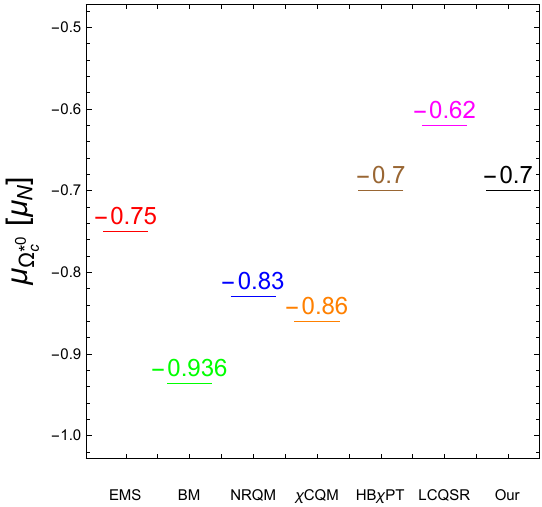}}
 \caption{Magnetic dipole moments of the $\rm{J^P} =\frac{3}{2}^+$ singly-charm sextet baryons obtained in different approaches.}
 \label{comp4}
  \end{figure}

\begin{figure}[htb!]
%\centering
\subfloat[]{\includegraphics[width=0.35\textwidth]{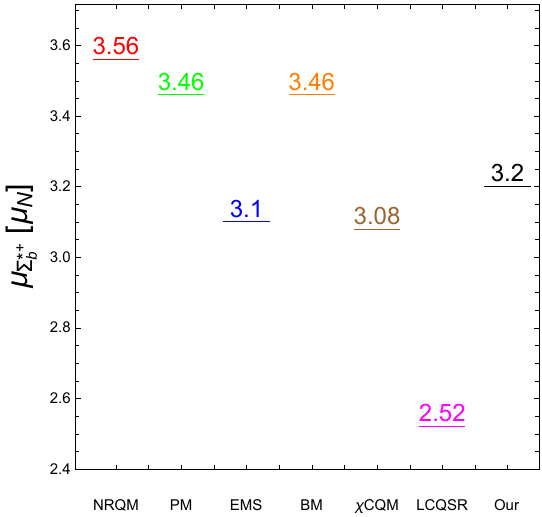}}~~~~~~~~~~
\subfloat[]{\includegraphics[width=0.35\textwidth]{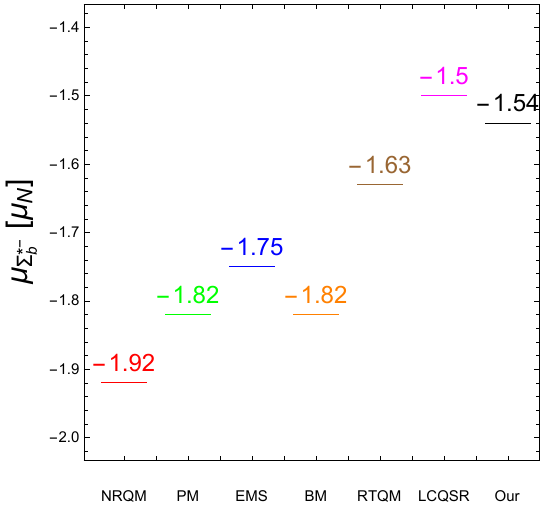}}\\
\subfloat[]{\includegraphics[width=0.35\textwidth]{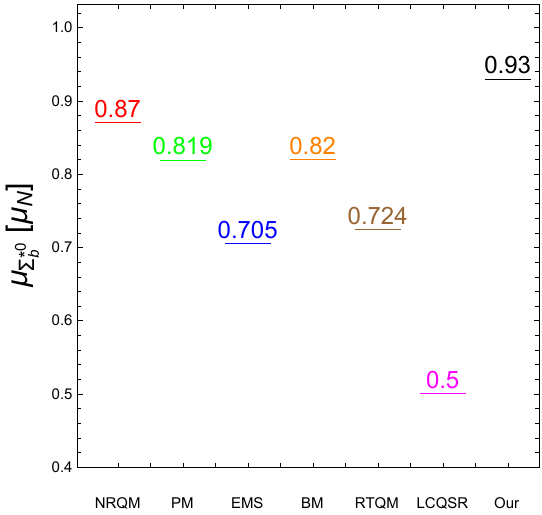}}~~~~~~~~~~
\subfloat[]{\includegraphics[width=0.35\textwidth]{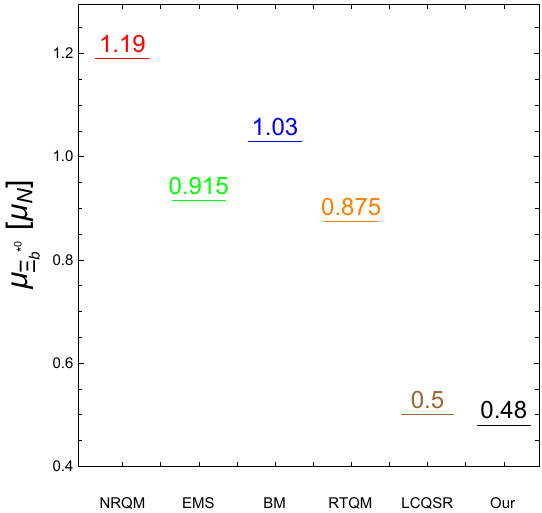}}\\
\subfloat[]{\includegraphics[width=0.35\textwidth]{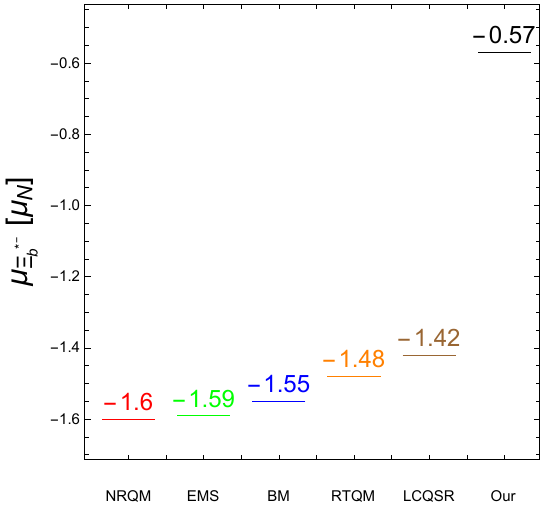}}~~~~~~~~~~
\subfloat[]{\includegraphics[width=0.35\textwidth]{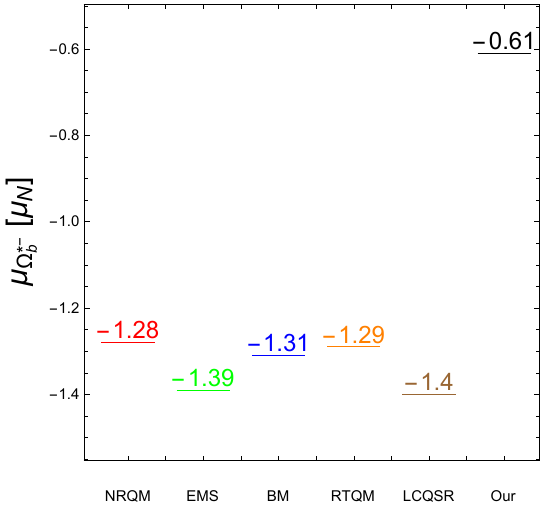}}
 \caption{Magnetic dipole moments of the $\rm{J^P} =\frac{3}{2}^+$ singly-bottom sextet baryons obtained in different approaches.}
 \label{comp5}
  \end{figure}

The numerical results obtained yielded the following observations: 

\begin{itemize}
 \item Upon analysis of the obtained results, it becomes evident that the short-distance interaction of quarks with the photon, i.e. the perturbative contribution, constitutes the predominant component of the analysis, accounting for approximately $(65-79)\%$. The remaining contributions are derived from the long-distance interactions of light quarks with the photon, which is a non-perturbative contribution.
 
 \item  A more profound and detailed insight into the dynamics of quarks can be attained through a comprehensive examination of the distinct contributions of the individual quark sector. Upon conclusion of this analysis, it is observed that the magnetic dipole moments of the spin-$\frac{1}{2}$ sextet singly-heavy baryons are subject to the influence of the light quarks. Conversely, the role of the heavy quark is significantly enhanced for the spin-$\frac{1}{2}$ anti-triplet and spin-$\frac{3}{2}$ sextet singly-heavy baryons.
 
 \item  The contribution of light- and heavy-quarks is observed to have an inverse relationship. The signs of the magnetic dipole moments demonstrate the interaction of the spin degrees of freedom of the quarks. The opposing signs of the light- and heavy-quark magnetic dipole moments indicate that their spins are anti-aligned in the baryon.
 
 \item To gain further insight, a comparative analysis between the numerical values obtained in this study and the existing literature on the subject is recommended. These can be found in Tables \ref{octet_charm_mm}-\ref{octet_bottom_mm2} and Figs. \ref{comp1}-\ref{comp5}.  The results obtained for $\rm{J^P}=\frac{1}{2}^+$ singly-charm sextet baryons are in general agreement with those obtained from lattice QCD and a few other models. However, the results obtained for $\rm{J^P}=\frac{1}{2}^+$ singly-charm anti-triplet baryons using different models are quite different. It can be seen that there is no consistency in the results for these baryons.  The results obtained for the $\rm{J^P}=\frac{3}{2}^+$ singly-charm sextet baryons are generally consistent with those of other models, with a few exceptions.  Upon analysis of singly-bottom baryons, it became evident that our findings have not aligned with those of other models, except a few models, particularly in the case of anti-triplet baryons. The discrepancy between the various theoretical approaches in predicting the magnetic dipole moments of singly-heavy baryons may be primarily attributed to the selection of wave functions and state mixing effects.  Nevertheless, the source of this evident discrepancy remains unresolved. Further theoretical and experimental studies are necessary to elucidate these inconsistencies and to gain a deeper comprehension of the current situation. However, direct measurements of the magnetic dipole moments of singly-heavy baryons are not yet feasible. Consequently, any indirect projections of the magnetic dipole moments of the singly-heavy baryons would be highly beneficial.
 
 \item As a final remark on the magnetic dipole moments, it is indeed possible to determine the ratio of the violation of the $U$-symmetry in the results obtained. The $U$-symmetry violation has been considered through a nonzero s-quark mass and different s-quark condensate.  Upon analysis,  it was observed that for spin-$\frac{1}{2}$ singly-heavy baryons, the  $U$-symmetry violation is reasonable ($\geq 20\%$),  with the exception of the $  \frac{\Sigma_c^0}{ \Omega_c^0}$ case.  In the case of the  spin-$\frac{3}{2}$ singly-heavy baryons,  it was observed that the  $U$-symmetry violation is large ($<20\%$), except for the $\frac{\Sigma_c^{*+}}{\Xi_c^{*+}}$ case. A similar scenario of a violation of the $U$-symmetry is observed in the results of alternative theoretical models of the magnetic dipole moments. % A comparable scenario of $U$-symmetry violation is evident in the findings of alternative theoretical models concerning the magnetic dipole moments.

 %Upon analysis, it was observed that for spin-$\frac{1}{2}$  and spin-$\frac{3}{2}$ baryons, the  $U$-symmetry violation is large ($>15\%$), with the exception of the $\frac{\Sigma_c^{+}}{\Xi_c^{\prime +}}$,  $\frac{\Sigma_b^{0}}{\Xi_b^{ \prime 0}}$, $  \frac{\Sigma_b^-}{ \Omega_b^-}$, $ \frac{\Lambda_b^0}{\Sigma_b^0}$ and $\frac{\Sigma_c^{*+}}{\Xi_c^{*+}}$ cases. A comparable scenario of large $U$-symmetry violation is evident in the findings of alternative theoretical models concerning the magnetic dipole moments.
 
 \item To guarantee the comprehensiveness of the analysis, the electric quadrupole and magnetic octupole moments of the $\rm{J^P}=\frac{3}{2}^+$ singly-heavy baryons are also calculated. The anticipated outcomes are enumerated in Table \ref{parameter3}. It can be observed that the magnitudes of the electric quadrupole and magnetic octupole moments are considerably less than that of the magnetic dipole moments. We ascertained the existence of non-zero values for the electric quadrupole and magnetic octupole moments of these baryons, indicative of a non-spherical charge distribution. The sign of electric quadrupole moment is positive for $\Sigma_c^{*++}$, $\Sigma_c^{*+}$, $\Xi_c^{*+}$, $\Xi_c^{*0}$,  $\Sigma_b^{*+}$, $\Sigma_b^{*0}$, $\Xi_b^{*0}$ and negative for $\Sigma_c^{*0}$, $\Omega_c^{*0}$, $\Sigma_b^{*-}$, $\Xi_b^{*-}$, $\Omega_b^{*-}$ which correspond to the prolate and oblate charge distributions, respectively. 
It is well established that the signs of the higher multipole moments provide information about the deformation of the associated baryon and its direction.  In the case of $\Sigma_c^{*++}$, $\Sigma_c^{*+}$, $\Xi_c^{*+}$, $\Xi_c^{*0}$,  $\Sigma_b^{*+}$, $\Sigma_b^{*-}$, and $\Xi_b^{*0}$ baryons, it is obtained that both the electric quadrupole and magnetic octupole moments have same sign and same geometric shape as the charge distribution. In the case of $\Sigma_c^{*0}$, $\Omega_c^{*0}$, $\Sigma_b^{*0}$, $\Xi_b^{*-}$, and $\Omega_b^{*-}$ baryons, it is obtained that opposite signs for the electric quadrupole and magnetic octupole moments. This indicates that the charge distribution and geometric shape are opposite for these baryons.

\end{itemize}

\section{summary and concluding notes}\label{summary}

The electromagnetic characteristics of singly-heavy baryons at low energies are responsive to their internal composition, structural configuration, and the associated chiral dynamics of light diquarks.  To gain further insight, experimentalists are attempting to measure the magnetic and electric dipole moments of charm baryons at the LHC. In view of these developments, we conducted an extensive analysis of the magnetic dipole moments of both      $\rm{J^P}=\frac{1}{2}^+$ and $\rm{J^P}=\frac{3}{2}^+$ singly-heavy baryons by means of the QCD light-cone sum rules. Our findings have been compared with other phenomenological estimations that could prove a valuable supplementary resource for interpreting the singly-heavy baryon sector. To shed light on the internal structure of these baryons we study the contributions of the individual quark sectors to the magnetic dipole moments. It was observed that the magnetic dipole moments of the spin-$\frac{1}{2}$ sextet singly-heavy baryons are governed by the light quarks. Conversely, the role of the heavy quark is significantly enhanced for the spin-$\frac{1}{2}$ anti-triplet and spin-$\frac{3}{2}$ sextet singly-heavy baryons. The contribution of light and heavy quarks is observed to have an inverse relationship. The signs of the magnetic dipole moments demonstrate the interaction of the spin degrees of freedom of the quarks. The opposing signs of the light and heavy-quark magnetic dipole moments imply that the spins of these quarks are anti-aligned with respect to each other in the baryon. As a byproduct, the electric quadrupole and magnetic octupole moments of spin-$\frac{3}{2}$ singly-heavy baryons are also calculated.  We ascertained the existence of non-zero values for the electric quadrupole and magnetic octupole moments of these baryons, indicative of a non-spherical charge distribution.

Future experimental endeavors pertaining to the properties of singly-heavy flavor baryons have the potential to elucidate discrepancies among disparate model predictions. It is our hope that our findings will prove instrumental in future experimental and theoretical pursuits pertaining to singly-heavy baryons.

 \section{Acknowledgments}
The author would like to acknowledge A. \"{O}zpineci for his invaluable contributions to the comments, discussions, and suggestions presented in this work.

\newpage

%\begin{widetext}
%
\begin{table}[htb!]
	\addtolength{\tabcolsep}{10pt}
	\caption{Working regions of $\rm{s_0}$ and $\rm{M^2}$ for the magnetic dipole moments of the singly-heavy baryons with $\rm{J^P} =\frac{1}{2}^+$.}
	\label{parameter}
	%	\begin{center}
	\begin{ruledtabular}
		%\scalebox{1.0}{
\begin{tabular}{|c|c|c|c|}
 %               \hline\hline
           %    \\
Baryons & $\mu \,[\mu_N]$ & $\rm{s_0}$ [GeV$^2$] & $\rm{M^2}$ [GeV$^2$] %&    PC ($\%$) %~~ %& ~~  CVG  ($\%$) 
\\
                \hline\hline                                        \hline
			\multicolumn{4}{|l|}{Sextet-C=1 }\\
			\hline
 $~ \Sigma_c^{++}$& $~~2.02 \pm 0.18$& $9.2-10.2$ & $2.0-2.6$ %& $51.18-72.95$ %& % $<2.0$  
                        \\
$  \Sigma_c^+$& $~~0.50 \pm 0.05$& $9.2-10.2$ & $2.0-2.6$ %& $51.72-69.96$ %&  %$<2.5$  
                       \\
$ \Sigma_c^0$& $-1.01 \pm 0.09$& $9.2-10.2$ & $2.0-2.6$ %& $51.88-72.50$ %&  %$<2.5$   
                       \\
$ ~\Xi_c^{\prime +}$& $~~0.49 \pm 0.05$ & $9.5-10.5$ & $2.4-3.0$ %& $56.45-72.72$ %&  %$<2.5$   
                       \\
$ ~\Xi_c^{\prime 0}$& $-0.68 \pm 0.06$ & $9.5-10.5$ & $2.3-2.7$ %& $51.09-72.80$ %&  %$<2.5$  
\\
$ \Omega_c^{ 0}$&$-0.73 \pm 0.08$ & $10.2-11.4$ & $2.4-3.0$ %& $52.98-70.05$ %&  %$<2.5$  
\\
\hline
			\multicolumn{4}{|l|}{Anti-triplet-C=1 }\\
			\hline
$ \Lambda_c^{+}$& $~~0.46 \pm 0.09$& $7.5-8.5$ & $1.4-1.8$ %& $54.43-68.51$ %&  %$<2.5$  
\\
$ \Xi_c^{+}$& $~~0.33 \pm 0.05$& $7.5-8.5$ & $1.4-1.8$ %& $50.46-64.15$ %&  %$<2.5$ 
\\
$ \Xi_c^{ 0}$&  $~~0.41 \pm 0.05$& $7.5-8.5$ & $1.4-1.8$ %& $53.27-70.37$ %& % $<2.5$ 
\\
\hline
\hline
			\multicolumn{4}{|l|}{Sextet-B=1 }\\
			\hline
$ \Sigma_b^{+}$&$~~2.02 \pm 0.19$ & $40.0-42.0$ & $5.2-6.2$ %& $54.44-70.37$ %&  %$<2.5$ 
\\
$ \Sigma_b^{0}$& $~~0.53 \pm 0.06$ & $40.0-42.0$ & $5.2-6.2$ %& $51.04-75.45$ %&  %$<2.5$ 
\\
$ \Sigma_b^{-}$& $-1.01 \pm 0.9$& $40.0-42.0$ & $5.2-6.2$%& $55.40-70.80$ %&  %$<2.5$ 
\\
$ \Xi_b^{\prime 0}$&$~~0.49 \pm 0.05$ & $41.0-43.0$ & $5.8-7.0$ %& $54.34-74.72$ %&  %$<2.5$   
                       \\
$ \Xi_b^{\prime -}$& $-0.78 \pm 0.10$ & $41.0-43.0$ & $5.5-6.5$ %& $51.54-67.52$ %&  %$<2.5$   
                       \\
$ \Omega_b^{-}$&$-0.87 \pm 0.07$ & $44.0-46.0$ & $3.0-3.6$ %& $53.75-68.06$ %&  %$<2.5$   
                       \\
                       \hline
			\multicolumn{4}{|l|}{Anti-triplet-B=1 }\\
			\hline
$ \Lambda_b^{0}$& $-0.29 \pm 0.03$& $40.0-42.0$ & $4.6-5.6$ %& $50.36-65.07$ %&  %$<2.5$  
\\
$ \Xi_b^{0}$&$-0.33 \pm 0.03$ & $40.0-42.0$ & $4.6-5.6$ %& $50.24-64.86$ %&  %$<2.5$ 
\\
$ \Xi_b^{-}$&$-0.25 \pm 0.03$ & $40.0-42.0$ & $4.5-5.5$ %& $51.51-66.65$ %&  %$<2.5$ 
\\
  %                                     \hline\hline
 \end{tabular}
%}
%\end{center}
\end{ruledtabular}
\end{table}
%%%%%%%%%%%%%%%%%%%%%%%%%%%%%%%%%%%%%%%%%%%%%%%%%%%%%%%%%%%
%\end{widetext}
 \begin{table}[htb!]
	\addtolength{\tabcolsep}{10pt}
	\caption{Working regions of $\rm{s_0}$ and $\rm{M^2}$ for the magnetic dipole moments of the singly-heavy baryons with $\rm{J^P} =\frac{3}{2}^+$.}
	\label{parameter2}
	%	\begin{center}
	\begin{ruledtabular}
		%\scalebox{1.0}{
\begin{tabular}{|c|c|c|c|}
 %               \hline\hline
           %    \\
Baryons & $\mu \,[\mu_N]$ & $\rm{s_0}$ [GeV$^2$] & $\rm{M^2}$ [GeV$^2$] %&    PC ($\%$) %~~ %& ~~  CVG  ($\%$) 
\\
                                        \hline\hline
                                        \hline
			\multicolumn{4}{|l|}{Sextet-C=1 }\\
			\hline
 $~ \Sigma_c^{*++}$& $~~3.40 \pm 0.34$& $9.5-10.5$ & $2.2-2.6$ %& $51.18-72.95$ %& % $<2.0$  
                        \\
$  \Sigma_c^{*+}$& $~~0.90 \pm 0.10$& $9.5-10.5$ & $2.2-2.6$ %& $51.72-69.96$ %&  %$<2.5$  
                       \\
$ \Sigma_c^{*0}$& $-1.44 \pm 0.19$& $9.5-10.5$ & $2.2-2.6$ %& $51.88-72.50$ %&  %$<2.5$   
                       \\
$ ~\Xi_c^{* +}$& $~~0.80 \pm 0.08$ & $9.8-11.0$ & $2.2-2.8$ %& $56.45-72.72$ %&  %$<2.5$   
                       \\
$ ~\Xi_c^{* 0}$& $-0.51 \pm 0.05$ & $9.8-11.0$ & $2.2-2.8$ %& $51.09-72.80$ %&  %$<2.5$  
\\
$ \Omega_c^{* 0}$&$-0.70 \pm 0.05$ & $11.5-13.0$ & $2.4-2.8$%&$52.98-70.05$ %&  %$<2.5$  
\\
\hline
\hline
\hline
			\multicolumn{4}{|l|}{Sextet-B=1 }\\
			\hline
$ \Sigma_b^{*+}$&$~~3.20 \pm 0.26$ & $41.0-43.0$ & $5.5-7.0$ %& $54.44-70.37$ %&  %$<2.5$ 
\\
$ \Sigma_b^{*0}$& $~~0.93 \pm 0.10$ & $41.0-43.0$ & $5.5-7.0$ %& $51.04-75.45$ %&  %$<2.5$ 
\\
$ \Sigma_b^{*-}$& $-1.54 \pm 0.12$& $41.0-43.0$ & $5.5-7.0$%& $55.40-70.80$ %&  %$<2.5$ 
\\
$ \Xi_b^{* 0}$&$~~0.48 \pm 0.05$ & $42.0-44.0$ & $5.5-6.5$ %& $54.34-74.72$ %&  %$<2.5$   
                       \\
$ \Xi_b^{* -}$& $-0.57 \pm 0.05$ & $42.0-44.0$ & $5.5-6.5$ %& $51.54-67.52$ %&  %$<2.5$   
                       \\
$ \Omega_b^{*-}$&$- 0.61 \pm 0.05$ & $44.5-46.5$ & $6.0-7.0$ %& $53.75-68.06$ %&  %$<2.5$   
                       \\
  %                                     \hline\hline
 \end{tabular}
%}
%\end{center}
\end{ruledtabular}
\end{table}
%%%%%%%%%%%%%%%%%%%%%%%%%%%%%%%%%%%%%%%%%%%%%%%%%%%%%%%%%%%%
%\end{widetext}
 \begin{table}[htb!]
	\addtolength{\tabcolsep}{10pt}
	\caption{The electric quadrupole $(\mathcal{Q})$ moments and magnetic octupole $(\mathcal{O})$ moments of the singly-heavy baryons with $\rm{J^P} =\frac{3}{2}^+$.}
	\label{parameter3}
	%	\begin{center}
	\begin{ruledtabular}
		%\scalebox{1.0}{
\begin{tabular}{|c|c|c|}
 %               \hline\hline
           %    \\
Baryons &  $\mathcal{Q} \times (10^{-2})$ [fm$^2$] & $\mathcal{O} \times (10^{-3})$ [fm$^3$] %&    PC ($\%$) %~~ %& ~~  CVG  ($\%$) 
\\
                                        \hline\hline
                                        \hline
			\multicolumn{3}{|l|}{Sextet-C=1 }\\
			\hline
 $~ \Sigma_c^{*++}$& $2.08 \pm 0.18$ & $1.35 \pm 0.13$ %& $51.18-72.95$ %& % $<2.0$  
                        \\
$  \Sigma_c^{*+}$& $0.81 \pm 0.08$ & $0.78 \pm 0.08$ %& $51.72-69.96$ %&  %$<2.5$  
                       \\
$ \Sigma_c^{*0}$& $-0.10 \pm 0.02$ & $0.37 \pm 0.05$ %& $51.88-72.50$ %&  %$<2.5$   
                       \\
$ ~\Xi_c^{* +}$ & $0.73 \pm 0.07$ & $0.68 \pm 0.06$ %& $56.45-72.72$ %&  %$<2.5$   
                       \\
$ ~\Xi_c^{* 0}$ & $0.26 \pm 0.04$ & $0.49 \pm 0.05$ %& $51.09-72.80$ %&  %$<2.5$  
\\
$ \Omega_c^{* 0}$& $-0.081 \pm 0.009$ & $0.20 \pm 0.03$%&$52.98-70.05$ %&  %$<2.5$  
\\
\hline
\hline
\hline
			\multicolumn{3}{|l|}{Sextet-B=1 }\\
			\hline
$ \Sigma_b^{*+}$ & $1.49 \pm 0.05$ & $0.21 \pm 0.02$ %& $54.44-70.37$ %&  %$<2.5$ 
\\
$ \Sigma_b^{*0}$ & $0.034 \pm 0.008$ & $-0.051 \pm 0.007$ %& $51.04-75.45$ %&  %$<2.5$ 
\\
$ \Sigma_b^{*-}$& $-1.00 \pm 0.05$ & $-0.23 \pm 0.02$%& $55.40-70.80$ %&  %$<2.5$ 
\\
$ \Xi_b^{* 0}$ & $0.48 \pm 0.03$ & $0.18 \pm 0.02$ %& $54.34-74.72$ %&  %$<2.5$   
                       \\
$ \Xi_b^{* -}$& $-0.053 \pm 0.011$ & $0.091 \pm 0.006$ %& $51.54-67.52$ %&  %$<2.5$   
                       \\
$ \Omega_b^{*-}$ & $-0.19 \pm 0.02$ & $0.019 \pm 0.003$ %& $53.75-68.06$ %&  %$<2.5$   
                       \\
  %                                     \hline\hline
 \end{tabular}
%}
%\end{center}
\end{ruledtabular}
\end{table}
%%%%%%%%%%%%%%%%%%%%%%%%%%%%%%%%%%%%%%%%%%%%%%%%%%%%%%%%%%%%%%%%%%%%
%\end{widetext}
	\begin{table}[htb!]
		\caption {Magnetic dipole moments of $\rm{J^P}=\frac{1}{2}^+$ singly-charm baryons in units of the nuclear magneton $\mu_N$.}
		\label{octet_charm_mm}
		%\footsize
		%\begin{threeparttable}
		\begin{ruledtabular}
		\begin{tabular}{|c|c|c|c|c|c|c|c|c|c|}	\hline  
			\textbf{Baryons} &\textbf{EMS} & \textbf{BM}  & \textbf{NRQM} & \textbf{RTQM} & \textbf{HB$\chi$PT} & \textbf{B$\chi$PT} & \textbf{LCQSR} & \textbf{LQCD}& \textbf{This Work}\\
			%\cline{1-2} 
			& \bf\cite{Mohan:2022sxm}& \bf\cite{Simonis:2018rld}& \bf\cite{Bernotas:2012nz}& \bf\cite{Faessler:2006ft} & \bf\cite{Wang:2018gpl}& \bf\cite{Shi:2018rhk, Liu:2018euh} & \bf\cite{Aliev:2015axa, Aliev:2001ig, Aliev:2008ay}& \bf\cite{Can:2021ehb,Bahtiyar:2016dom}&\\   \hline	\hline
			\multicolumn{10}{|l|}{Sextet-C=1 }\\
			\hline
$\Sigma_c^{++}$   & $2.095$   & $2.280$    & $2.350$  & $1.760$  & $1.50$  & $2.00$  & $2.40$ & $2.220(505)$ & $~~2.02(18)$ \\	
$\Sigma_c^{+}$    & $0.432$  & $0.487$  &  $0.490$  & $0.360$  & $0.30$  & $0.46$  & $0.50$    & $-$& $~~0.50(5)$          \\	
$\Sigma_c^{0}$    & $-1.234$  &$-1.310$   & $-1.370$ & $-1.040$ & $-0.91$ & $-1.08$ & $-1.50$   & $-1.073 (269)$ & $-1.01(9)$ \\
$\Xi_c^{\prime+}$ & $0.623$   & $0.633$    & $0.890$  & $0.470$  & $0.31$ & $0.62$  & $0.80$    & $0.315(141)$ & $~~0.49(5)$    \\
$\Xi_c^{\prime0}$ & $-1.074$  & $-1.120$  & $-1.180$ & $-0.950$ & $-0.80$ & $-0.91$ & $-1.20$   & $-0.599(71)$ & $-0.68(6)$    \\   	
$\Omega_c^{0}$    & $-0.905$ & $-0.950$ &  $-0.940$ & $-0.850$ & $-0.69$ & $-0.74$ & $-0.90$   & $-0.639(88)$ & $-0.73(8)$   \\ \hline
			\multicolumn{10}{|l|}{Anti-triplet-C=1 }\\\hline
$\Lambda_c^{+}$   & $0.38$     & $0.335$    & $0.39$  & $0.42$  & $0.24$  & $0.24$  & $0.40(5)$  & $-$  & $0.46(9)$            \\
$\Xi_c^{+}$  & $0.38$    & $0.334$    & $0.20$  & $0.41$  & $0.29$  & $0.24$  & $0.50(5)$  & $0.235(25)$ & $0.33(5)$             \\
$\Xi_c^{0}$  & $0.38$   & $0.334$    & $0.41$  & $0.39$  & $0.19$  & $0.19$  & $0.35(5)$  & $0.192(17)$  & $0.41(5)$            \\\hline	
		\end{tabular}
		%\end{threeparttable}
		\end{ruledtabular}
	\end{table}
	%%%%%%%%%%%%%%%%%%%%%%%%%%%%%%%%%%%%%%%%%%%%%%%%%%%%%%%%%%%%%%%%
 \begin{table}[htb!]
		\caption {Magnetic dipole moments of $\rm{J^P}=\frac{1}{2}^+$ singly-bottom baryons in units of the nuclear magneton $\mu_N$.}
		\label{octet_bottom_mm}
		%\footsize
		%\begin{threeparttable}
		\begin{ruledtabular}
		\begin{tabular}{|c|c|c|c|c|c|c|c|}	\hline  
			\textbf{Baryons}&\textbf{NRQM}&\textbf{PM} &\textbf{EMS} & \textbf{BM}   & \textbf{RTQM} & \textbf{LCQSR} & \textbf{This Work}\\
			%\cline{1-2} 
			&\bf\cite{Bernotas:2012nz}&\bf\cite{Majethiya:2008fe}& \bf\cite{Dhir:2013nka}& \bf\cite{Simonis:2018rld}& \bf\cite{Faessler:2006ft}  & \bf\cite{Aliev:2015axa, Aliev:2001ig, Aliev:2008ay}& \\   \hline	\hline
			\multicolumn{8}{|l|}{Sextet-B=1 }\\
			\hline
$\Sigma_b^{+}$ & 2.50 & 2.12  & $2.19$   & $2.25$    &  $2.07$    & $2.40$     & $~~2.02(19) $ \\	
$\Sigma_b^{0}$ & 0.64 & 0.547   & $0.563$  &$0.603$   & $0.53$ & $0.60$    & $~~0.53(6)$ \\
$\Sigma_b^{-}$ &$-1.22$ &$-1.03$   & $-1.064$  & $-1.15$    & $-1.01$   & $-1.30$    & $-1.01(9)$             \\	
$\Xi_b^{\prime 0}$ & 0.90 & 0.658 & $0.756$   & $0.782$     & $0.66$   & $0.70$     & $~~0.49(5)$    \\
$\Xi_b^{\prime -}$ &$-1.02$&$-0.941$ & $-0.913$  & $-0.968$   & $-0.91$ & $-1.20$   & $-0.78(10)$    \\   	
$\Omega_b^{-}$ &$-0.79$ &$-0.805$  & $-0.741$ & $-0.806$ &   $-0.82$ &  $-0.80$    & $-0.87(7)$   \\ \hline
			\multicolumn{8}{|l|}{Anti-triplet-B=1 }\\\hline
$\Lambda_b^0$ &$-0.066$&$-0.064$  & $-0.062$     & $-0.06$     & $-0.06$  & $-0.18(5)$  & $-0.29(3)$            \\
$\Xi_b^{0}$ &$-0.11$ & $-$     & $-0.062$    & $-0.06$     & $-0.06$  & $-0.08(2)$   & $-0.33(3)$             \\
$\Xi_b^{-}$ &$-0.05$ &$-$     & $-0.062$   & $-0.0555$      & $-0.06$    & $-0.045(5)$    & $-0.26(3)$            \\\hline	
		\end{tabular}
		%\end{threeparttable}
		\end{ruledtabular}
	\end{table}
%%%%%%%%%%%%%%%%%%%%%%%%%%%%%%%%%%%%%%%%%%%%%%%%%%%%%%%%%%%%%%%%%%%%%%% 
 	\begin{table}[htb!]
		\caption {Magnetic dipole moments of $\rm{J^P}=\frac{3}{2}^+$ singly-charm baryons in units of the nuclear magneton $\mu_N$.}
		\label{octet_charm_mm2}
		%\footsize
		%\begin{threeparttable}
		\begin{ruledtabular}
		\begin{tabular}{|c|c|c|c|c|c|c|c|c|}	\hline  
			\textbf{Baryons} &\textbf{EMS} & \textbf{BM}  & \textbf{NRQM} & \textbf{$\chi$CQM} & \textbf{HB$\chi$PT}  & \textbf{LCQSR} & \textbf{LQCD}& \textbf{This Work}\\
			%\cline{1-2} 
			& \bf\cite{Mohan:2022sxm}& \bf\cite{Simonis:2018rld}& \bf\cite{Bernotas:2012nz}& \bf\cite{Sharma:2010vv} & \bf\cite{Wang:2018gpl}& \bf\cite{Aliev:2008sk}& \bf\cite{Can:2021ehb}&\\   \hline	\hline
			\multicolumn{9}{|l|}{Sextet-C=1 }\\
			\hline
$\Sigma_c^{*++}$   & $3.578$   & $3.98$    & $4.11$  & $3.92$  & $2.41$  & $4.81(122)$   & $-$ & $~~3.40(34)$ \\	
$\Sigma_c^{*+}$    & $1.185$  & $1.337$  &  $1.32$  & $0.97$  & $0.67$    & $2.00(46)$    & $-$ & $~~0.90(10)$          \\	
$\Sigma_c^{*0}$    & $-1.214$  &$-1.49$   & $-1.47$ & $-1.99$ & $-1.07$ & $-0.81(20)$    & $-$ & $-1.44(19)$ \\
$\Xi_c^{*+}$ & $1.453$      & $1.47$  & $1.64$  & $1.59$ & $0.81$  & $1.68(42)$    & $-$ & $~~0.80(8)$    \\
$\Xi_c^{*0}$ & $-0.987$    & $-1.2$ & $-1.15$ & $-1.43$ & $-0.90$ & $-0.68(18)$   &$-$ & $-0.51(5)$    \\   	
$\Omega_c^{*0}$     & $-0.75$ &  $-0.936$ & $-0.83$ & $-0.86$ & $-0.70$ & $-0.70(18)$   & $-0.730(23)$ & $-0.70(5)$   \\ \hline
					\end{tabular}
		%\end{threeparttable}
		\end{ruledtabular}
	\end{table}
%%%%%%%%%%%%%%%%%%%%%%%%%%%%%%%%%%%%%%%%%%%%%%%%%%%%%%%%%%%%%%%%%%%5
	 \begin{table}[htb!]
		\caption {Magnetic dipole moments of $\rm{J^P}=\frac{3}{2}^+$ singly-bottom baryons in units of the nuclear magneton $\mu_N$.}
		\label{octet_bottom_mm2}
		%\footsize
		%\begin{threeparttable}
		\begin{ruledtabular}
		\begin{tabular}{|c|c|c|c|c|c|c|c|}	\hline  
\textbf{Baryons}&\textbf{NRQM}&\textbf{PM} &\textbf{EMS} & \textbf{BM}   & \textbf{$\chi$CQM} & \textbf{LCQSR} & \textbf{This Work}\\
			%\cline{1-2} 
&\bf\cite{Bernotas:2012nz}&\bf\cite{Majethiya:2008fe}& \bf\cite{Dhir:2013nka}& \bf\cite{Simonis:2018rld}& \bf\cite{Sharma:2010vv}  & \bf\cite{Aliev:2008sk}& \\   \hline	\hline
			\multicolumn{8}{|l|}{Sextet-B=1 }\\
			\hline
$\Sigma_b^{*+}$ &3.56&3.46  &3.10  & 3.46    &  3.08    & 2.52(50)     & $~~3.20(26) $ \\	
$\Sigma_b^{*0}$ &0.87&0.819   & $0.705$  &$0.82$   & $0.724$ & $0.50(15)$    & $~~0.93(10)$ \\
$\Sigma_b^{*-}$ &$-1.92$ &$-1.82$  & $-1.75$  & $-1.82$    & $-1.63$   & $-1.50(36)$    & $-1.54(12)$            
\\	
$\Xi_b^{* 0}$ &1.19 &$-$ & 0.915   & 1.03    & 0.875   & 0.50(15)     & $~~0.48(5)$    \\
$\Xi_b^{* -}$ &$-1.60$&$-$ & $-1.59 $ & $-1.55$  & $-1.48$ & $ -1.42(35) $  & $-0.57(5)$    \\   	
$\Omega_b^{*-}$ &$-1.28$ & $-$ &$-1.39$ & $-1.31$ &  $ -1.29$ & $-1.40(35)$    & $-0.61(5)$   \\ \hline
					\end{tabular}
		%\end{threeparttable}
		\end{ruledtabular}
	\end{table}
%\begin{widetext}
%
%\begin{figure}[htb!]
%\centering
%\subfloat[]{\includegraphics[width=0.47\textwidth]{1-DSigma_cMsq.pdf}}~~~~~~~~~~
%\subfloat[]{\includegraphics[width=0.47\textwidth]{1-DSigma_cs0.pdf}}\\
%\subfloat[]{\includegraphics[width=0.47\textwidth]{1-DSigmacstarMsq.pdf}}~~~~~~~~~~
%\subfloat[]{\includegraphics[width=0.47\textwidth]{1-DSigmacstars0.pdf}}\\
%\subfloat[]{\includegraphics[width=0.47\textwidth]{1-DstarSigmacMsq.pdf}}~~~~~~~~~~
%\subfloat[]{\includegraphics[width=0.47\textwidth]{1-DstarSigmacs0.pdf}}
% \caption{The magnetic dipole moments of the singly-heavy baryons versus $\rm{M^2}$ (left panel), and $\rm{s_0}$ (right panel).}
% \label{Msqfig1}
%  \end{figure}
%
%  \end{widetext}

  \newpage
  
\bibliographystyle{elsarticle-num}
\bibliography{SinglyBaryonMM.bib}

\end{document}